\documentclass[aps,prd,floatfix,superscriptaddress,nofootinbib,12pt]{revtex4-1}
\usepackage[utf8]{inputenc}
\usepackage{epsfig,latexsym,cancel,amssymb,amsmath,verbatim,mathrsfs}
\usepackage{xcolor}
\usepackage{subfigure}
\usepackage{graphicx}
\usepackage{bm}
\usepackage{multirow}
\allowdisplaybreaks[4]

\definecolor{ccblue}{rgb}{0.0,0.4,0.8}
\usepackage{hyperref}
\hypersetup{colorlinks=true,
  linkcolor=ccblue,
  urlcolor=ccblue,
  citecolor=ccblue}

\usepackage{tikz}
\usetikzlibrary{arrows,shapes}
\usetikzlibrary{trees}
\usetikzlibrary{matrix,arrows} 				
\usetikzlibrary{positioning}				
\usetikzlibrary{calc,through}				
\usetikzlibrary{decorations.pathreplacing}  
\usepackage{pgffor}							
\usetikzlibrary{decorations.pathmorphing}	
\usetikzlibrary{decorations.markings}
\tikzset{
    vector/.style={decorate, decoration={snake}, draw},
	provector/.style={decorate, decoration={snake,amplitude=2.5pt}, draw},
	antivector/.style={decorate, decoration={snake,amplitude=-2.5pt}, draw},
    fermion/.style={draw=black, postaction={decorate},
        decoration={markings,mark=at position .55 with {\arrow[draw=black]{>}}}},
    fermionbar/.style={draw=black, postaction={decorate},
        decoration={markings,mark=at position .55 with {\arrow[draw=black]{<}}}},
    fermionnoarrow/.style={draw=black},
    gluon/.style={decorate, draw=black,
        decoration={coil,amplitude=4pt, segment length=5pt}},
    scalar/.style={dashed,draw=black, postaction={decorate},
        decoration={markings,mark=at position .55 with {\arrow[draw=black]{>}}}},
    scalarbar/.style={dashed,draw=black, postaction={decorate},
        decoration={markings,mark=at position .55 with {\arrow[draw=black]{<}}}},
    scalarnoarrow/.style={dashed,draw=black},
    electron/.style={draw=black, postaction={decorate},
        decoration={markings,mark=at position .55 with {\arrow[draw=black]{>}}}},
	bigvector/.style={decorate, decoration={snake,amplitude=4pt}, draw},
}
\tikzstyle{block} = [draw, rectangle, 
    minimum height=3em, minimum width=6em]

\begin{document}

\title{Combined explanations of $B$-physics anomalies,  $(g-2)_{e, \mu}$ and neutrino masses by scalar leptoquarks}
\author{Shao-Long Chen}
\email[E-mail: ]{chensl@mail.ccnu.edu.cn}
\affiliation{Key Laboratory of Quark and Lepton Physics (MoE) and Institute of Particle Physics, Central China Normal University, Wuhan 430079, China}
\affiliation{Center for High Energy Physics, Peking University, Beijing 100871, China}

\author{Wen-wen Jiang}
\email[E-Mail: ]{wwjiang@mails.ccnu.edu.cn}
\affiliation{Key Laboratory of Quark and Lepton Physics (MoE) and Institute of Particle Physics, Central China Normal University, Wuhan 430079, China}

\author{Ze-Kun Liu}
\email[E-mail: ]{zekunliu@mails.ccnu.edu.cn}
\affiliation{Key Laboratory of Quark and Lepton Physics (MoE) and Institute of Particle Physics, Central China Normal University, Wuhan 430079, China}

\begin{abstract}
We extend the contents of the standard model (SM) by introducing TeV-scale scalar leptoquarks to generate neutrino masses and explain some current observed deviations from the SM predictions, including the anomalous magnetic moments of charged leptons (electron and muon)  and $B$-physics anomalies ($R_{K^{(*)}}$ and $R_{D^{(*)}}$). The model consists of $\text{SU}(2)_L$ singlet leptoquark $S_1\sim (\bar{3}, 1, 1/3)$, doublet leptoquark $\tilde{R}_2\sim (3, 2, 1/6)$ and triplet leptoquark $S_3\sim (\bar{3}, 3, 1/3)$. We combine the constraints arising from the low-energy lepton flavor violation, meson decay and mixing observables.  We perform a detailed phenomenological analysis and identify the minimized texture of leptoquark Yukawa matrices to accommodate a unified explanation of the anomalies and neutrino oscillation data.
\end{abstract}

\maketitle

\section{Introduction}
The neutrino oscillation experiments have firmly established that neutrinos are massive and have non-trivial mixing between different generations~\cite{Super-Kamiokande:1998kpq, SNO:2002tuh, KamLAND:2002uet, T2K:2011ypd}. The experiments also indicate that the neutrino masses are much smaller than that of charged fermions, which suggests that neutrinos may have specific sources of mass generation. In the recent decades, a plethora of models have been proposed to explain the neutrino mass and the natural way is the so called seesaw mechanism~\cite{Ma:1998dn}. Type-I seasaw model~\cite{Minkowski:1977sc, Mohapatra:1979ia, Gell-Mann:1979vob, Glashow:1979nm, Schechter:1980gr} provides neutrino masses at the tree-level by extending the particle content of the SM with three $\text{SU}(2)_L$-singlet right-handed neutrino fields, while type-II~\cite{Schechter:1980gr, Mohapatra:1980yp, Lazarides:1980nt} and type-III~\cite{Foot:1988aq} models introduce  $\text{SU}(2)_L$-triplet scalar and $\text{SU}(2)_L$-triplet fermions, respectively. Beyond tree level, the tiny neutrino masses could radiatively originate from loop levels~\cite{Zee:1980ai, Cheng:1980qt, Zee:1985id, Babu:1988ki, Cai:2017jrq}.

Extending the SM to include the source of the origin of neutrino mass and mixing brings new physics, especially to the flavor sector. The intensity frontier precision measurements may pin down the possible connections between neutrino physics and flavor physics. Such as the anomalous magnetic moments of electron and muon, there are long-standing discrepancies between the theoretical predictions and measured values~\cite{Aoyama:2017uqe, Keshavarzi:2018mgv, Davier:2019can, Aoyama:2020ynm,Aoyama:2012wk,Aoyama:2019ryr,Czarnecki:2002nt,Gnendiger:2013pva,Davier:2017zfy,Keshavarzi:2018mgv,Colangelo:2018mtw,Hoferichter:2019mqg,Davier:2019can,Keshavarzi:2019abf,Kurz:2014wya,Melnikov:2003xd,Masjuan:2017tvw,Colangelo:2017fiz,Hoferichter:2018kwz,Gerardin:2019vio,Bijnens:2019ghy,Colangelo:2019uex,Blum:2019ugy,Colangelo:2014qya,Abi:2021gix}. The anomalies also include the ratios $R_{K^{(*)}}$ and $R_{D^{(*)}}$ in $B$-decays, pointing towards the lepton flavor universality violation, measured by BaBar~\cite{BaBar:2012obs, BaBar:2013mob}, Belle~\cite{Belle:2015qfa, Belle:2016dyj, Belle:2016kgw} and LHCb~\cite{LHCb:2017vlu, LHCb:2017smo, LHCb:2019hip, LHCb:2017avl, LHCb:2021trn} collaborations. In this work, we propose a model with scalar leptoquarks to provide a common explanation of  neutrino mass and these flavor anomalies.

Leptoquarks (LQs) have been introduced in many new physics models beyond the SM and are very popular to explain $B$-physics anomalies with one or more leptoquark states~\cite{Dorsner:2016wpm, Crivellin:2020mjs, Carvunis:2021dss}. The unified solution to both $R_{K^{(*)}}$ and $R_{D^{(*)}}$ anomalies seems rule out single scalar leptoquark models~\cite{Angelescu:2018tyl}. Among the scalar leptoquarks, triplet $S_3\sim (\bar{3}, 3, 1/3)$ can accommodate the $R_{K^{(*)}}$ anomalies, while the $R_{D^{(*)}}$ anomalies can be resolved by introducing either a singlet $S_1\sim (\bar{3}, 1, 1/3)$ or a doublet $R_2\sim (3, 2, 7/6)$ leptoquark. The double leptoquarks models were proposed to explain both $R_{K^{(*)}}$ and $R_{D^{(*)}}$ anomalies, involving $S_1$  and $S_3$ combination~\cite{Bigaran:2019bqv, Saad:2020ihm, Gherardi:2020qhc, Greljo:2021xmg, Lee:2021jdr, Bhaskar:2022vgk} or $R_2$ and $S_3$ combination~\cite{Chen:2016dip, Saad:2020ucl, Babu:2020hun}. Extending with leptoquarks will give contribution to  the anomalous magnetic moment of charged lepton at one-loop level and the no-chiral scalar leptoquarks  $S_1$ or $R_2$, which have both left-chiral and right-chiral couplings, can provide good explanations to the $a_{\mu}$ and $a_{e}$ deviations~\cite{Dorsner:2020aaz, Bigaran:2020jil} simultaneously. The mixing between different type leptoquarks can also generate non-trivial Majorana neutrino mass terms at one-loop level. The minimal model to generate neutrino mass by the scalar leptoquark mixing requires a pair of leptoquarks and the possible combinations are $S_1-\tilde{R}_2 (3, 2, 1/6)$, $S_3-\tilde{R}_2$ and $S_3 - R_2$~\cite{AristizabalSierra:2007nf, Dorsner:2017wwn, Julio:2022ton, Julio:2022bue, Chowdhury:2022dps}. Motivated by the leptoquark abundant phenomenologies, we attempt to extend the SM contents by scalar leptoquarks to generate neutrino mass and explain the flavor anomalies mentioned above. 

This paper is organized as follow: In Section II, we briefly introduce the model set-up and the neutrino mass generation mechanism. In Section III, we show how to explain the flavor anomalies in the model, including $R_{K^{(*)}}$,  $R_{D^{(*)}}$, $a_{\mu}$ and $a_{e}$. We discuss the observables constraints on the leptoquark couplings in Section IV and then we perform a detailed analysis of model parameter space and identify two benchmark points in Section V and we conclude in the final section.

\section{The model and neutrino mass generation}
\subsection{The model}
In addition to the SM fields, we introduce three scalar leptoquarks, including an $\text{SU}(2)_L$ singlet $S_1\sim (\bar{3}, 1, 1/3)$, a doublet $\tilde{R}_2\sim (3, 2, 1/6)$ and a triplet $S_3\sim (\bar{3}, 3, 1/3)$. The scalar leptoquarks are denoted as
\begin{align}
&S_1(\bar{3}, 1, 1/3)=S_1^{1/3}\,, \quad
\tilde{R}_2(3, 2, 1/6)= (\tilde{R}_2^{2/3}, \tilde{R}_2^{-1/3})^{\text{T}}\,, \nonumber \\[0.2cm]
& S_3(\bar{3}, 3, 1/3) =  \tau^i S_3^i = \begin{pmatrix}
S_3^{1/3} & \sqrt{2}\,S_3^{4/3} \\
\sqrt{2}\,S_3^{-2/3} & -S_3^{1/3}
\end{pmatrix}\,,
\end{align}
where $\tau^i$ ($i=1, 2, 3$) are the Pauli matrices and we define $S_3^{4/3}=(S_3^1-\text{i} S_3^2)/\sqrt{2}$, $S_3^{-2/3}=(S_3^1+\text{i} S_3^2)/\sqrt{2}$ and $S_3^{1/3}=S_3^3$. The corresponding Yukawa terms that describe the interactions between leptoquarks and fermions are given by
\begin{align}
\mathcal{L}_{Y} =&-y_{1R}^{ij} \overline{u^{iC}_R} e^j_R S_1 - y_{1L}^{ij}\overline{Q_L^{iC}}\, i\tau^2 \, L_L^j S_1 -y_{2L}^{ij}\ \overline{d^{i}_R} \tilde{R}_2^{T} \, i\tau^2 \, L_L^j \nonumber\\
& - y_{3L}^{ij}\ \overline{Q_L^{iC}}\, i\tau^2 \, S_3 L_L^j + \text{h.c.}\,,
\label{Yukawa}
\end{align}
where $Q$ and $L$ denote the $\text{SU}(2)_L$ doublet left-handed quarks and leptons, $u_R$, $d_R$ and $e_R$ denote the $\text{SU}(2)_L$ singlet right-handed up-type quarks, down-type quarks and charged leptons,  respectively. All fields in Eq.~(\ref{Yukawa}) are represented in the flavor basis. For phenomenological analysis, it is more convenient that we re-parametrize the couplings in the fermion mass basis. The Yukawa coupling terms are then rewritten in the mass basis of fermions as the following form,
\begin{align}
\mathcal{L}_{Y} =& -y_{1R}^{ij} \overline{u^{iC}_R} e^j_R S_1^{1/3} + (V^Ty_{1L})^{ij}\overline{d^{iC}_L} \nu_L^j S_1^{1/3} - y_{1L}^{ij}\overline{u^{iC}_L} e_L^j S_1^{1/3} + y_{2L}^{ij}\ \overline{d^i_R} \nu^j_L \tilde{R}_2^{-1/3}  \nonumber\\
&-y_{2L}^{ij}\ \overline{d^i_R} e_L^j \tilde{R}_2^{2/3} + (V^Ty_{3L})^{ij}\ \overline{d^{iC}_L} \nu^j_L S_3^{1/3}+ \sqrt{2} (V^Ty_{3L})^{ij}\ \overline{d^{iC}_L} e^j_L S_3^{4/3}\nonumber \\
&- \sqrt{2} y_{3L}^{ij}\ \overline{u^{iC}_L} \nu^j_L S_3^{-2/3} + y_{3L}^{ij}\ \overline{u^{iC}_L} e^j_L S_3^{1/3} + \text{h.c.}\,.
\end{align}
where $V$ is the CKM matrix.  Since in our analysis of $(g-2)_{e,\mu}$ and $B$-physics anomalies, the choice of neutrino mass or flavor basis has negligible effect, the neutrino states in the above equation are kept in flavor basis.

The renormalizable and gauge invariant scalar potential involving $H$, $S_1$, $\tilde{R}_2$ and $S_3$ is described by
\begin{align}
V \supset \,& m_{H}^2\, H^{\dagger} H + m_{1}^2\, S_{1}^{\dagger} S_{1} + m_{2}^2\, \tilde{R}_{2}^{\dagger} \tilde{R}_{2} + \frac{1}{2}m_{3}^2\, \text{Tr}\big(S_3^{\dagger} S_3\big) + \lambda_H\left(H^{\dagger} H\right)^2 + \lambda_1 \left(S_1^{\dagger} S_1\right)^2  \nonumber\\
&+ \lambda_{2} \left(\tilde{R}_2^{\dagger} \tilde{R}_2\right)^2 + \lambda_3\left[\text{Tr}(S_3^{\dagger} S_3)\right]^2 + \lambda_3^{\prime}\,\text{Tr}(S_3^{\dagger} S_3^{\dagger})\text{Tr}(S_3 S_3) + \lambda_{H1}\, H^{\dagger} H S_1^{\dagger} S_1 \nonumber\\
&+ \lambda_{H2}\, H^{\dagger} H \tilde{R}_2^{\dagger} \tilde{R}_2 + \frac{1}{2} \lambda_{H3}\, H^{\dagger} H\,\text{Tr}\big(S_3^{\dagger} S_3\big) + \big(\lambda_{13}\, H^{\dagger}S_3^{\dagger}HS_1 + \mu_1\, \tilde{R}_2^{\dagger} H S_1^* \nonumber\\
&+ \mu_2\, \tilde{R}_2^{\dagger} S_3^{\dagger} H + \text{h.c.}\big)\,,
\end{align}
where $H$ is the SM Higgs doublet. More general interactions of leptoquarks and SM Higgs can be found in Ref.~\cite{Dorsner:2022twk}. After the spontaneous electroweak symmetry breaking, the Higgs field $H$ acquires a vacuum expecting value (VEV) with $\langle H \rangle=v/\sqrt2, v=246\,\text{GeV}$. The physical scalar particles include one electric neutral Higgs boson $h$, three $1/3$-charged leptoquarks, two $2/3$-charged leptoquarks, and one $4/3$-charged leptoquark. In the basis of $\rho^{1/3}\equiv (S_1^{1/3}, \tilde{R}_{2}^{1/3}, S_3^{1/3})^{T}$ and $\rho^{2/3}\equiv (\tilde{R}_{2}^{2/3}, S_3^{2/3})^{T}$, the mass matrices for the two groups of charged scalar particles are given by
\begin{align}
&M_{1/3}^2= \begin{pmatrix}
m_{1}^2+\frac{1}{2}\lambda_{H1}v^2 & \frac{1}{\sqrt{2}} \mu_1 v & -\frac{1}{2}\lambda_{13} v^2 \\[0.2cm]
\frac{1}{\sqrt{2}}\mu_1 v         & m_{2}^2 + \frac{1}{2}\lambda_{H2}v^2 &  -\frac{1}{\sqrt{2}} \mu_2 v \label{massm1} \\[0.2cm]
-\frac{1}{2}\lambda_{13} v^2 &  -\frac{1}{\sqrt{2}} \mu_2 v  & m_3^2  + \frac{1}{2}\lambda_{H3}v^2 
\end{pmatrix}\,, \\[0.31cm]
&M_{2/3}^2= \begin{pmatrix}
m_{2}^2 + \frac{1}{2}\lambda_{H2}v^2     & \mu_2 v\\[0.2cm]
\mu_2 v                       & m_3^2  + \frac{1}{2}\lambda_{H3}v^2 
\end{pmatrix}\,. \label{massm2}
\end{align}
 After diagonalization of the above mass matrices, we obtain the physical scalar fields: charge-1/3 leptoquarks $(\phi_1, \phi_2, \phi_3)$ and charge-2/3 leptoquarks $(\omega_1, \omega_2)$, which satisfy
\begin{align}
&\phi_i=R^{1/3}_{ij}\,\rho^{1/3}_j\,, \\
&\omega_i=R^{2/3}_{ij}\,\rho^{2/3}_j\,,
\end{align}
where $R^{1/3}$ and $R^{2/3}$ are the corresponding rotation matrices. The rotation matrix $R^{2/3}$ can be  parametrized as
\begin{equation}
R^{2/3}=\begin{pmatrix}
\cos \alpha  & \sin \alpha \\
-\sin \alpha & \cos \alpha
\end{pmatrix}\,,
\end{equation}
where the mixing angle is given by
\begin{equation}
\tan 2 \alpha = \frac{2 \mu_{2}v}{m_{2}^{2}-m_{3}^{2}+(\lambda_{H2}-\lambda_{H3})v^{2}/2 }\,.
\end{equation}
The rotation matrix $R^{1/3}$ need three rotation angles to be parametrized,
\begin{equation}
R^{1/3}=R(\theta_{12})R(\theta_{13})R(\theta_{23})\,.
\end{equation}
In the limit where off-diagonal elements are much smaller than the diagonal elements, the mixing angle in the rotation matrix $R^{1/3}$ can be approximatively calculated by
\begin{equation}
\theta_{ij} \simeq \frac{\left(M_{1/3}^2\right)_{ij}}{\left(M_{1/3}^2\right)_{ii}-\left(M_{1/3}^2\right)_{jj}}\,.
\label{estn}
\end{equation}
The charge-4/3 component $S_3^{4/3}$ has no mixing with other scalar fields and we denote the mass by $m_{S_3}$. In our analysis of the low energy processes, we assume the leptoquark multiplets to be quasi-degenerate and set the LQ masses as $m_{S_1} = m_{\phi_1}$, $m_{R_2} = m_{\phi_2}\approx m_{\omega_1}$ and $m_{S_3} = m_{\phi_3}\approx m_{\omega_2}$.
\subsection{Neutrino masses}
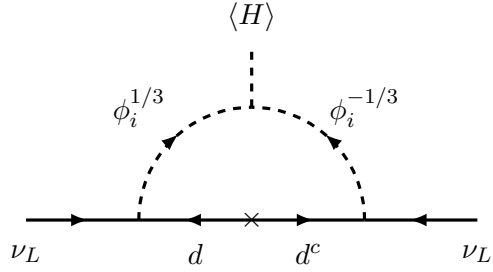
\begin{figure}[!h]
\centering
\begin{tikzpicture}[line width=1.2 pt, scale=1.5, >=latex]
	\draw[fermion] (-1,0)--(0,0);
	\draw[fermionbar] (0,0)--(1,0);
	\draw[scalar] (0,0) arc (180:90:1);
	\draw[scalar] (2,0) arc (0:90:1);
	\draw[scalarnoarrow] (1,1)--(1,1.5);
	\draw[fermion] (1,0)--(2,0);
	\draw[fermionbar] (2,0)--(3,0);
	\node at (1,0) {$\times$};
	\node at (-1,-0.3) {$\nu_{L}$};
	\node at (0.5,-0.3) {$d$};
	\node at (1.5,-0.3) {$d^c$};
	\node at (3,-0.3) {$\nu_{L}$};
	\node at (0,1) {$\phi^{1/3}_i$};
	\node at (2,1) {$\phi^{-1/3}_i$};
	\node at (1,1.8) {$\langle H \rangle$};
\end{tikzpicture}
\caption{Feynman diagram of Majorana neutrino masses generation at one-loop level.}
\label{NM-loop}
\end{figure}
In our model, the neutrino masses are induced at one-loop level through the Feynman diagrams as shown in Fig.~\ref{NM-loop}, in which the loop is mediated by the down-type quarks and 1/3-charged leptoquarks. The neutrino mass matrix is given by~\cite{AristizabalSierra:2007nf}
\begin{align}
({\cal M}_{\nu})_{\alpha\beta} =& \frac{3}{16\pi^2}\sum_{\substack{i=1,2,3 \\ k=d,s,b}} m_k B_0(0,m_k^2,m_{\phi_i}^2) \left\{R_{i1}^{1/3} R_{i2}^{1/3} \left[(V^Ty_{1L})^{k\alpha} y_{2L}^{k\beta}+ (V^Ty_{1L})^{k\beta} y_{2L}^{k\alpha}\right] \right. \nonumber \\
&\left.  + R_{i2}^{1/3} R_{i3}^{1/3}  \left[(V^Ty_{3L})^{k\alpha} y_{2L}^{k\beta}+ (V^Ty_{3L})^{k\beta} y_{2L}^{k\alpha}\right] \right\}\,,
\label{neutrinomass}
\end{align}
where $B_0(0,m_k^2,m_{\phi_i}^2)$ is the Passarino-Veltman function and its finite part is given by
\begin{equation}
B_0(0,m_k^2,m_{\phi_i}^2)=\frac{m_k^2 \log (m_k^2)-m_{\phi_i}^2 \log (m_{\phi_i}^2)}{m_k^2-m_{\phi_i}^2}\,.
\end{equation}
The first term in the bracket of Eq.~(\ref{neutrinomass}) is associated with the $S_1-\tilde{R}_2$ combination, while the second term is associated with the $S_3-\tilde{R}_2$ combination. To simplify the analysis, we consider one term dominates the other. For example, when $S_1-\tilde{R}_2$ contribution is dominant ($\mu_1 \gg \mu_2, \lambda_{13}v$), the neutrino mass matrix can be written as
\begin{equation}
({\cal M}_{\nu})_{\alpha\beta}=\big(\hat{y}_{1L}^T \Lambda y_{2L} + y_{2L}^T \Lambda^T \hat{y}_{1L} \big)_{\alpha\beta}\,,
\label{NMmatrix}
\end{equation}
where we define $\hat{y}_{1L} \equiv (V^Ty_{1L})$ and
\begin{equation}
\Lambda \equiv \begin{pmatrix}
\Lambda_d & 0 & 0 \\
0 & \Lambda_s & 0 \\
0 & 0 & \Lambda_b \\
\end{pmatrix}\,, \qquad \text{with} \ 
\Lambda_k\simeq\frac{3}{32\pi^2} m_k\frac{\sqrt{2}\mu_1 v}{m_{\phi_1}^2-m_{\phi_2}^2} \log\left( \frac{m_{\phi_1}^2}{m_{\phi_2}^2}\right)\,.
\label{lambdak}
\end{equation}
Using the method of master parametrization~\cite{Cordero-Carrion:2018xre, Cordero-Carrion:2019qtu}, we parametrize the coupling matrices $\hat{y}_{1L}$ and $y_{2L}$ as
\begin{align}
&\hat{y}_{1L}=\frac{1}{\sqrt{2}} \Sigma^{-1/2} W A \hat{D}^{1/2}U^{\dagger}\,, \label{YCM1}\\
&y_{2L}=\frac{1}{\sqrt{2}} \Sigma^{-1/2} W^* B \hat{D}^{1/2}U^{\dagger}\,,
\label{YCM2}
\end{align}
where $U$ is the $3 \times 3$ unitary neutrino mixing matrix, which brings the neutrino mass matrix to diagonal form by
\begin{equation}
U^T\, \mathcal{M}_{\nu}\, U=\text{diag}(m_1, m_2, m_3)\,.
\label{Dsqtm}
\end{equation}
The forms of matrix $\Sigma , W, A, B$ and $\hat{D}$ depend on the ranks of neutrino mass matrix $\mathcal{M}_{\nu}$ and matrix $\Lambda$. Neutrino oscillation data requires that $\cal{M}_{\nu}$ should contain two or three non-vanishing eigenvalues. In our numerical analysis of neutrino masses, for simplicity, we neglect the $d-$quark contribution in neutrino mass loop $(\Lambda_d=0)$ and consider the normal ordering neutrino mass hierarchy with $m_1=0$. In this scenario, the ranks of matrices $\mathcal{M}_{\nu}$ and $\Lambda$ are both 2 and the neutrino mass matrix $\mathcal{M}_{\nu}$ only depends on the second and third columns of couplings $\hat{y}_{1L}$ and $y_{2L}$. In this case, $\Sigma$ takes form as $\text{diag}(\Lambda_{s}, \Lambda_{b})$ and $\hat{D}$ takes the form as $\text{diag}(\kappa, m_{2}, m_{3})$, where $\kappa$ can be arbitrary value, since it can always be absorbed by rescaling relevant elements in matrices A and B. Eq.~(\ref{YCM1}) and Eq.~(\ref{YCM2}) give the elements of second and third columns of couplings $\hat{y}_{1L}$ and $y_{2L}$.  The matrix $W$ is a $2 \times 2$ unitary complex matrix that contains 4 real degrees of freedom. The matrices $A$ and $B$ are defined as $A=TC_1$ and $B=(T^T)^{-1}(C_1C_2+KC_2)$, where $T$ is an upper-triangular $2\times 2$ complex matrix with positive real values in the diagonal and contains 4 degrees of freedom, $K$ is a $2\times 2$ anti-symmetric complex matrix that contains 2 degrees of freedom. The matrices $C_1$ and $C_2$ are given by
\begin{equation}
C_1= \begin{pmatrix}
z_1\, & 0\ & 0\ \\
z_2\, & 0\ & 0\ 
\end{pmatrix}\,,\ \ \ \ 
C_2= \begin{pmatrix}
-1\ & 0\ & 0\ \\
0\  & 1\ & 0\ \\
0\  & 0\ & 1\ 
\end{pmatrix}\,,
\end{equation}
where $z_1$ and $z_2$ are two complex numbers that contains 2 degrees of freedom with the condition $z_1^2+z_2^2=0$. The possible values of the second and third columns of matrices $\hat{y}_{1L}$ and $y_{2L}$ can be obtained by scanning these 12 real free parameters.
\subsection{Effective Lagrangians}
The tree-level contributions of leptoquarks to the related phenomenologies can be described by the following effective Lagrangians,
\begin{align}
  \mathcal{L}_{\bar{q} q \bar{\ell} \ell} = -\frac{4G_F}{\sqrt{2}}
  \Big[ & \left(g^{LL}_{V,q}\right)^{ij,mn}(\bar q_L^i \gamma^\mu  q_L^j)(\bar \ell_L^m \gamma_\mu \ell_L^n) + \left(g^{RL}_{V,q}\right)^{ij,mn}(\bar q_R^i \gamma^\mu  q_R^j)(\bar \ell_L^m \gamma_\mu \ell_L^n) \nonumber \\
&+\left(g^{LR}_{V,q}\right)^{ij,mn}(\bar q_L^i \gamma^\mu  q_L^j)(\bar \ell_R^m \gamma_\mu \ell_R^n) 
+\left(g^{RR}_{V,q}\right)^{ij,mn}(\bar q_R^i \gamma^\mu  q_R^j)(\bar \ell_R^m \gamma_\mu \ell_R^n)\nonumber \\
&+ \left(g^{LL}_{S,q}\right)^{ij,mn}(\bar q_R^i  q_L^j) (\bar \ell_R^m \ell_L^n) +\left(g^{RR}_{S,q}\right)^{ij,mn}(\bar q_L^i  q_R^j) (\bar \ell_L^m \ell_R^n) \nonumber \\
&+ \left(g^{LL}_{T,q}\right)^{ij,mn}(\bar q_R^i \sigma^{\mu\nu} q_L^j) (\bar \ell_R^m \sigma_{\mu\nu} \ell_L^n) + \left(g^{RR}_{T,q}\right)^{ij,mn}(\bar q_L^i \sigma^{\mu\nu} q_R^j) (\bar \ell_L^m\sigma_{\mu\nu} \ell_R^n) \Big] \,,\label{EFL1}\\
 \mathcal{L}_{\bar{q} q \bar{\nu} \nu} =   \frac{4G_F}{\sqrt{2}}
\Big[& \left(h^{LL}_{V,q}\right)^{ij,mn}(\bar q_L^i \gamma^\mu  q_L^j)(\bar \nu_L^m \gamma_\mu \nu_L^n)+ \left(h^{RL}_{V,q}\right)^{ij,mn}(\bar q_R^i \gamma^\mu q_R^j)(\bar \nu_L^m \gamma_\mu \nu_L^n)\Big]\,, \label{EFL2}\\
  \mathcal{L}_{\bar{u} d \bar{\ell} \nu} = -\frac{4G_F}{\sqrt{2}}
  \Big[& \left(c^{LL}_{V}\right)^{ij,mn}(\bar u_L^i \gamma^\mu  d_L^j)(\bar \ell_L^m \gamma_\mu \nu_L^n) +\left(c^{LL}_{S}\right)^{ij,mn}(\bar u_R^i  d_L^j) (\bar \ell_R^m \nu_L^n) \nonumber \\
  &+ \left(c^{LL}_{T}\right)^{ij,mn}(\bar u_R^i \sigma^{\mu\nu} d_L^j) (\bar \ell_R^m \sigma_{\mu\nu} \nu^n_L) \Big]+ \text{h.c.}\,.
  \label{EFL3}
\end{align}
The Wilson coefficients at the leptoquark mass scale are determined by the combinations of Yukawa couplings and summarized in Table~\ref{LQWilson}. To analyze the low-energy processes, these Wilson coefficients are needed to RGE run down to the appropriate scale. We take the low-energy scale at the bottom-quark mass ($m_{b}=4.18\ \text{GeV}$) and the Wilson coefficients at the leading logarithm approximation can be calculated by the following form~\cite{Chetyrkin:1997dh, Gracey:2000am, Dorsner:2013tla, Hiller:2016kry},
\begin{equation}
C_J(\mu=m_b)=\left[ \frac{\alpha_s(m_b)}{\alpha_s(m_t)} \right]^{-\gamma_1^J/\beta^{(5)}_1} \left[ \frac{\alpha_s(m_t)}{\alpha_s(m_{\text{LQ}})} \right]^{-\gamma_1^J/\beta^{(6)}_1} C_J(\mu=m_{\text{LQ}})\,,
\end{equation}
with the QCD running coefficient $\beta^{(n_f)}_1=(2n_f-33)/6$, where $n_f$ is the relevant number of quark flavors at the hadronic scale. The coefficients $\gamma_1^J$ are the anomalous dimension and given by $\gamma_1^V=0$, $\gamma_1^S=2$ and $\gamma_1^T=-2/3$. In our numerical analysis, we use the package \textsl{Wilson}~\cite{Aebischer:2018bkb} to calculate the running of Wilson coefficients and obtain the following Wilson coefficient correlations between the two scales,
\begin{align}
c_V^{LL}(m_b) &= 1.01\, c_V^{LL}(m_{\text{LQ}})\,, \notag \\
\begin{pmatrix}
  c_S^{LL}(m_b) \\ c_T^{LL}(m_b)
\end{pmatrix} & =
\begin{pmatrix}
  1.64 & -0.275 \\
  -3.87 \times 10^{-3} & 0.867
\end{pmatrix}
\begin{pmatrix}
  c_S^{LL}(m_{\text{LQ}}) \\ c_T^{LL}(m_{\text{LQ}})
\end{pmatrix}
\end{align}
where we have taken the leptoquark masses scale as $m_{\text{LQ}}=1\ \text{TeV}$. In the following discussion of the various physical processes, we utilize the Flavio package~\cite{Straub:2018kue} to get the favored region of Wilson coefficients at the leptoquark mass scale.
\begin{table}[ht]
\begin{footnotesize}
\centering
 \begin{tabular}{|c|l|l|l|}\hline
        & \multicolumn{1}{|c|}{$S_1$} & \multicolumn{1}{|c|}{$\tilde{R}_2$} & \multicolumn{1}{|c|}{$S_3$} \\ \hline
 \multirow{6}{*}{$u^i \to u^j \bar{\ell}^{m} \ell^{n}$} & $g^{LL}_{V,u}=-(y_{1L})^{im}(y_{1L}^*)^{jn}$ & & $g^{LL}_{V,u}=-(y_{3L})^{im}(y_{3L}^*)^{jn}$ \\
 & $g^{RR}_{V,u}=-(y_{1R})^{im}(y_{1R}^*)^{jn}$& & \\ 
 & $g^{LL}_{S,u}=(y_{1R})^{im}(y_{1L}^*)^{jn}$ & & \\ 
 & $g^{RR}_{S,u}=(y_{1L})^{im}(y_{1R}^*)^{jn}$ & & \\ 
 & $g^{LL}_{T,u}=-\frac{1}{4}(y_{1R})^{im}(y_{1L}^*)^{jn}$ & & \\ 
 & $g^{RR}_{T,u}=-\frac{1}{4}(y_{1L})^{im}(y_{1R}^*)^{jn}$ & & \\ \hline
 $d^i \to d^j \bar{\ell}^{m} \ell^{n}$ & & $g^{RL}_{V,d}=(y_{2L})^{jm}(y_{2L}^*)^{in}$ & $g^{LL}_{V,d}=-2(V^Ty_{3L})^{im}(V^{\dagger}y_{3L}^*)^{jn}$ \\ \hline
 $u^i \to u^j \bar{\nu}^{m} \nu^{n}$ & & & $h^{LL}_{V,u}=2(y_{3L})^{im}(y_{3L}^*)^{jn}$ \\ \hline
 $d^i \to d^j \bar{\nu}^{m} \nu^{n}$ & $h^{LL}_{V,d}=-(V^Ty_{1L})^{im}(V^{\dagger}y_{1L}^*)^{jn}$ & $h^{RL}_{V,d}=-(y_{2L})^{jm}(y_{2L}^*)^{in}$ & $h^{LL}_{V,d}=(V^Ty_{3L})^{im}(V^{\dagger}y_{3L}^*)^{jn}$ \\ \hline
 \multirow{3}{*}{$d^i \to u^j \ell^{n} \bar{\nu}^{m}$} & $c^{LL}_{V}=(V^Ty_{1L})^{im}(y_{1L}^*)^{jn}$ & & $c^{LL}_{V}=-(V^Ty_{3L})^{im}(y_{3L}^*)^{jn}$ \\
 & $c^{LL}_{S}=-(V^Ty_{1L})^{im}(y_{1R}^*)^{jn}$ &  & \\
 & $c^{LL}_{T}=\frac{1}{4}(V^Ty_{1L})^{im}(y_{1R}^*)^{jn}$ & &  \\ \hline
 \end{tabular}
 \end{footnotesize}
  \caption{The corresponding Wilson coefficients (in units of $v^2/4\,m_{\text{LQ}}^2$) in Eqs.~(\ref{EFL1}, \ref{EFL2}, \ref{EFL3}) induced by the leptoquarks at tree level. The matching scale is set at the leptoquark mass scale.}
   \label{LQWilson}
\end{table}

\section{The flavor anomalies}
In this section, we present the observed flavor anomalies between current experimental observations and the SM predictions, and explore how to alleviate these tensions by introducing scalar leptoquarks in our framework.

\subsection{$R_K$ and $R_{K^*}$}
The first observed anomalies we consider are the lepton flavor universality violation ratios $R_K$ and $R_{K^*}$, which are defined as
\begin{equation} 
R_{K}=\frac{\text{Br}(B^+ \to K^+ \mu^+ \mu^-)}{\text{Br}(B^+ \to K^+ e^+ e^-)}\,, \quad
R_{K^{*}}=\frac{\text{Br}(B^0 \to K^{*0} \mu^+ \mu^-)}{\text{Br}(B^0 \to K^{*0} e^+ e^-)}\,.
\label{RK-df}
\end{equation}
The SM predictions~\cite{Bobeth:2007dw, Bordone:2016gaq} for these two ratios are 
\begin{equation}
R_K^{\rm{SM}}=1.0003\pm 0.0001\,,  \quad R_{K^*}^{\rm{SM}}=1.00 \pm 0.01\,.
\end{equation}
The new measurements of $R_K$ and $R_{K^*}$ at low $q^2$ region $[1.1,\, 6.0]\ \rm{GeV^2}$ by LHCb are given by~\cite{LHCb:2019hip, LHCb:2021trn}
\begin{equation}
R_K^{\rm{LHCb}}=0.846^{+0.042\,+0.013}_{-0.039\, -0.012}\,, \quad R_{K^*}^{\rm{LHCb}}=0.685^{+0.113}_{-0.069}\pm 0.047\,,
\end{equation}
which both give deviation larger than $2.5\, \sigma$  from the SM prediction values. These two processes are determined by the neutral current, $b \to s \ell^+ \ell^{-}$. The effective Hamiltonian relevant to our model can be described by~\cite{Buchalla:1995vs},
\begin{equation}
\mathcal{H}_{\text{eff}} = -\frac{4G_F}{\sqrt{2}} V_{tb}V_{ts}^* \Big[ \sum_{X=9, 10}(C_X^{\ell\ell} \mathcal{O}_X^{\ell\ell}+C_{X^{\prime}}^{\ell\ell} \mathcal{O}_{X^{\prime}}^{\ell\ell}) \Big] +\text{h.c.}\,,
\end{equation}
where the $C_{X}^{\ell\ell}$ and $C_{X^{\prime}}^{\ell\ell}$ denote the Wilson coefficients and $\mathcal{O}_X^{\ell\ell}$ and $\mathcal{O}_{X^{\prime}}^{\ell\ell}$ are the corresponding effective operators, which take form as
\begin{align}
&\mathcal{O}_9^{\ell\ell}=\frac{e^2}{(4 \pi)^2}(\bar{s}\gamma^{\mu} P_L b)(\bar{\ell}\gamma_{\mu}\ell)\,,\quad \mathcal{O}_{10}^{\ell\ell}=\frac{e^2}{(4 \pi)^2}(\bar{s}\gamma^{\mu} P_L b)(\bar{\ell}\gamma_{\mu}\gamma_5 \ell)\,, \nonumber \\
&\mathcal{O}_{9^{\prime}}^{\ell\ell}=\frac{e^2}{(4 \pi)^2}(\bar{s}\gamma^{\mu} P_R b)(\bar{\ell}\gamma_{\mu}\ell)\,,\quad \mathcal{O}_{10^{\prime}}^{\ell\ell}=\frac{e^2}{(4 \pi)^2}(\bar{s}\gamma^{\mu} P_R b)(\bar{\ell}\gamma_{\mu}\gamma_5 \ell)\,.
\label{Wilson910}
\end{align}
\begin{figure}[!t]
  \begin{minipage}{0.49\linewidth}
    \centerline{\includegraphics[width=1\textwidth]{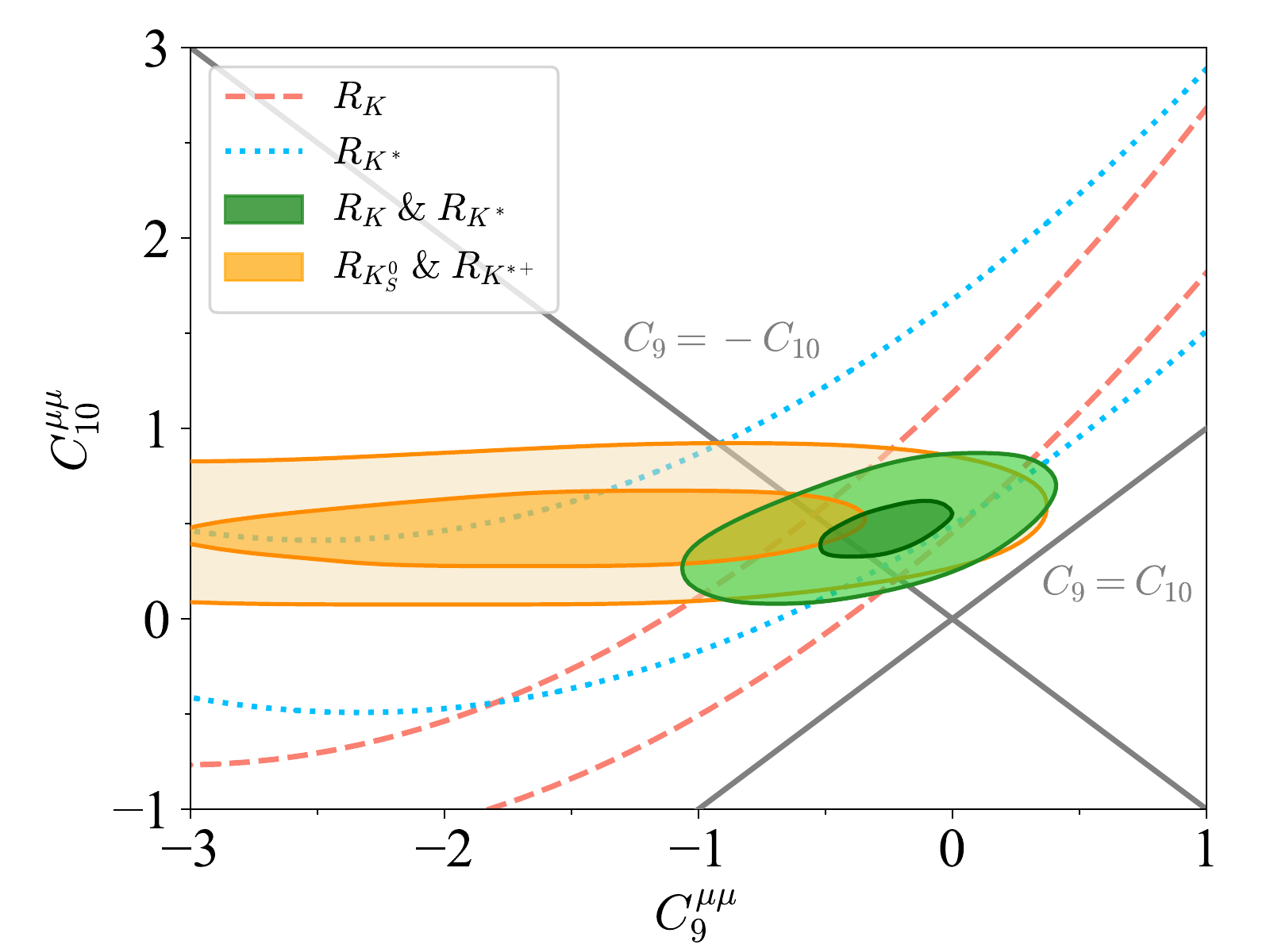}}
  \end{minipage}
  \begin{minipage}{0.49\linewidth}
    \centerline{\includegraphics[width=1\textwidth]{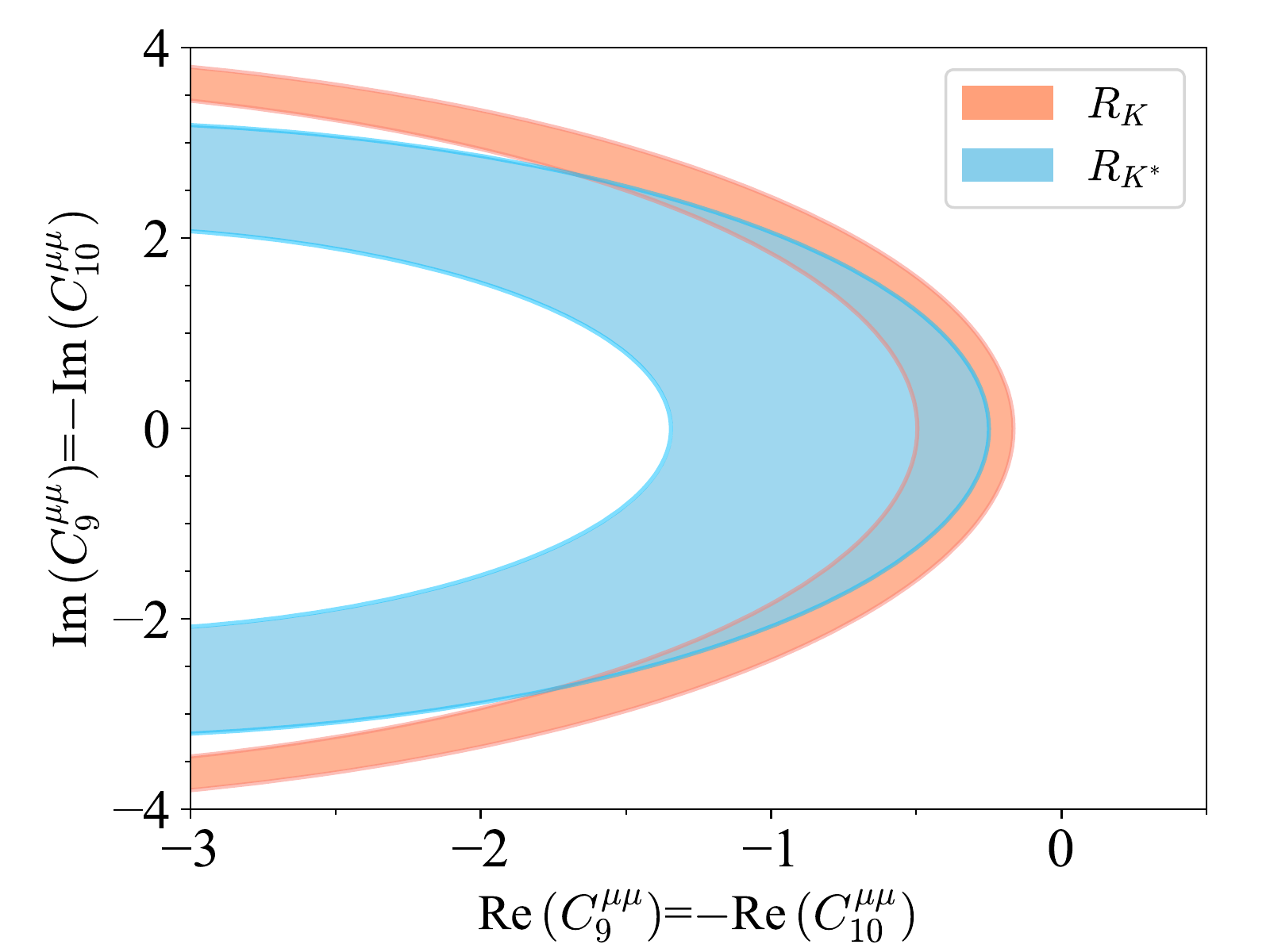}}
  \end{minipage}
    \begin{minipage}{0.49\linewidth}
    \centerline{\includegraphics[width=1\textwidth]{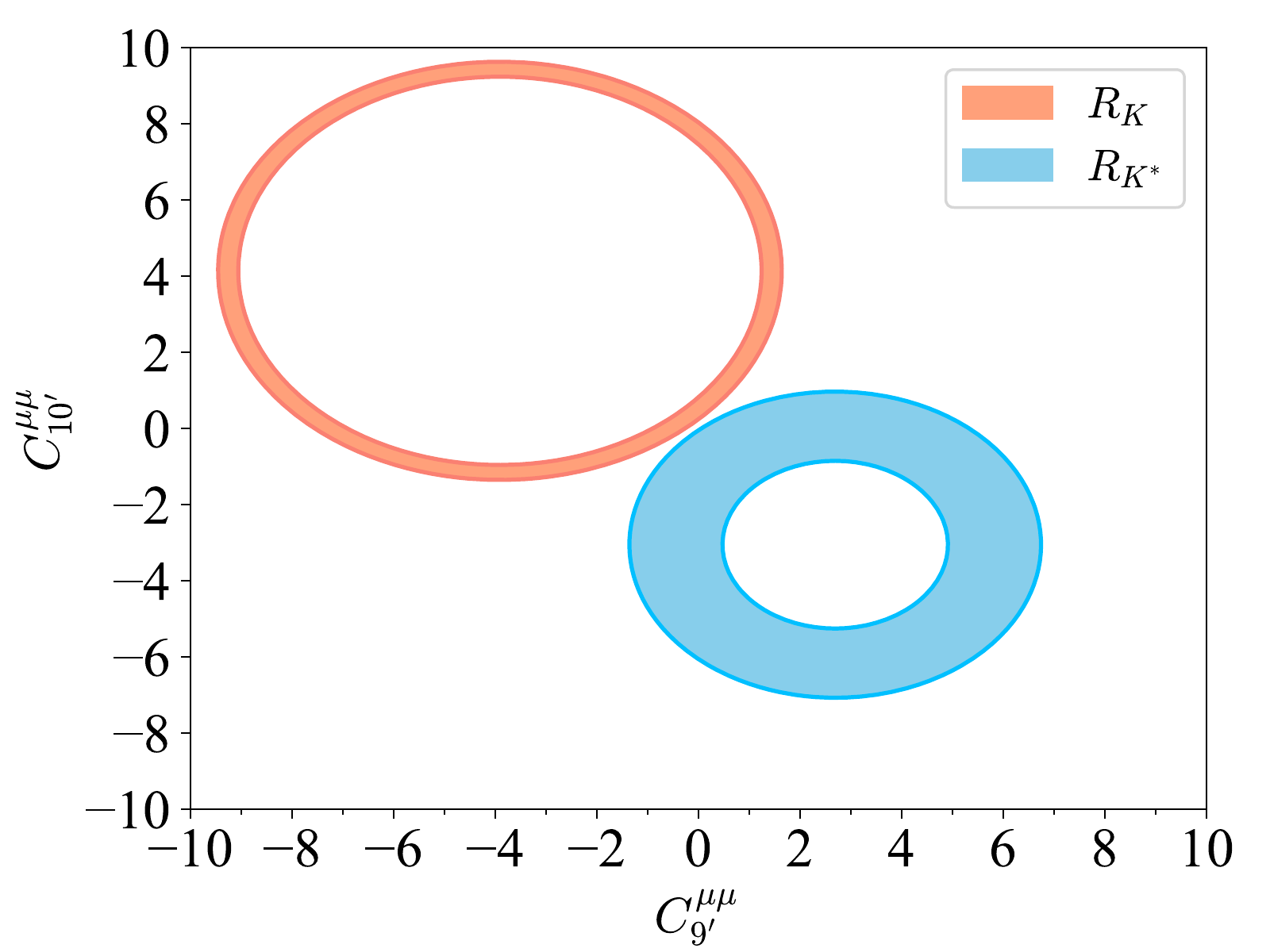}}
  \end{minipage}
  \begin{minipage}{0.49\linewidth}
    \centerline{\includegraphics[width=1\textwidth]{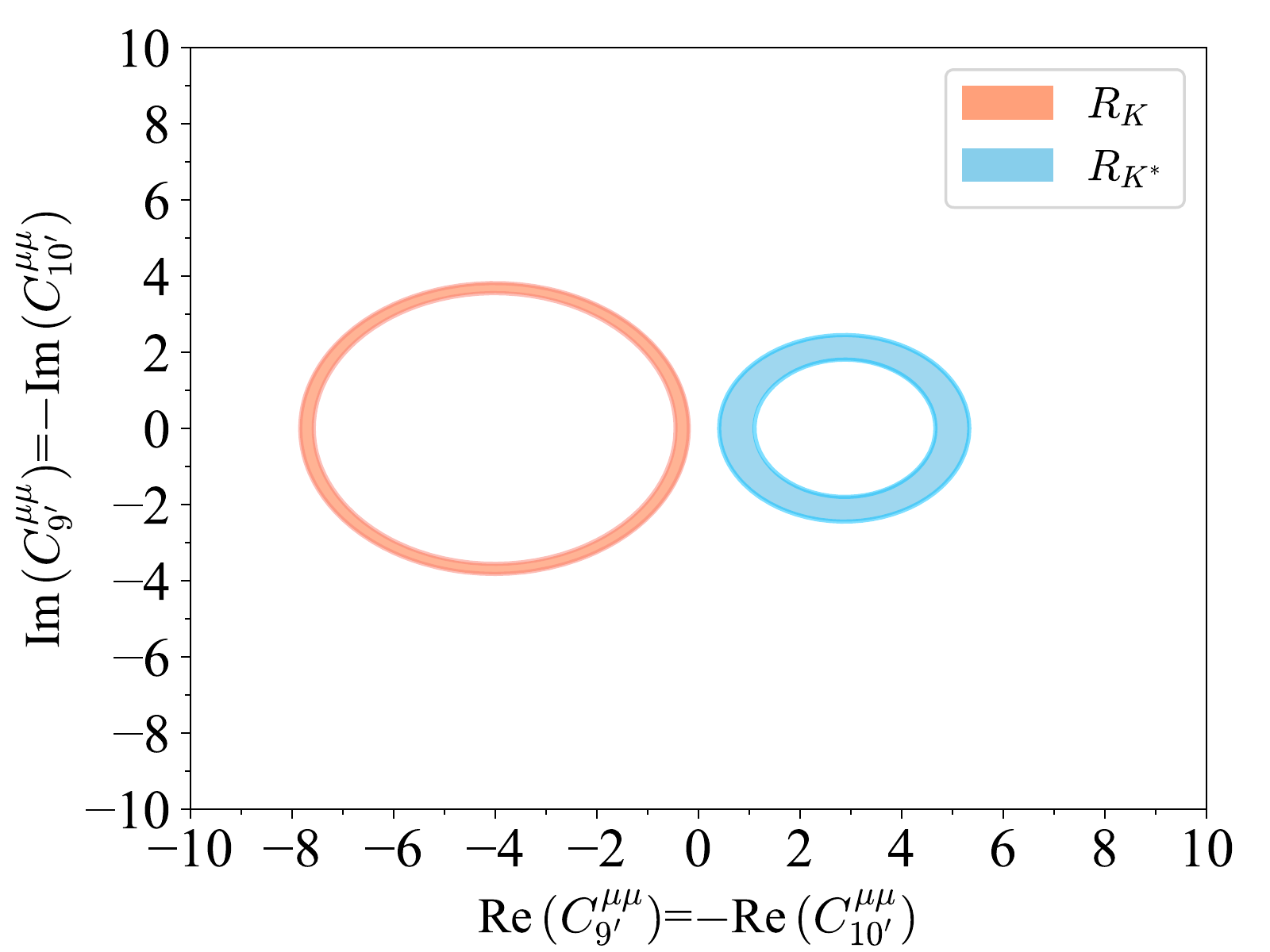}}
  \end{minipage}
  \caption{Contour plot of the fit to $R_K$ and $R_{K^*}$ in the plane of Wilson coefficients at the scale $\mu = 1\,\text{TeV}$. Upper left panel corresponds to $C_9^{\mu\mu}$ versus $C_{10}^{\mu\mu}$ as real. The dashed orange and dotted blue contours represent the $1\sigma$ allowed regions that explain $R_K$ and $R_{K^*}$ respectively. The dark green (yellow) region corresponds to the $1\sigma$ allowed region that explains $R_K$ and $R_{K^*}$ ($R_{K^0_S}$ and $R_{K^{*+}}$) simultaneously, while lighter region corresponds to $2\sigma$ allowed region. Upper right: The plot corresponds to the complex plane of $C_{9,10}^{\mu\mu}$, with $C_{9}^{\mu\mu}=- C_{10}^{\mu\mu}$ assumed. The orange and blue regions represent the $1\sigma$ allowed regions that explain $R_K$ and $R_{K^*}$, respectively. Bottom: Left panel shows the fit to $R_{K^{(*)}}$ using $C_{9^{\prime}}^{\mu\mu}$ and $C_{10^{\prime}}^{\mu\mu}$ as real parameters, while right one corresponds to the fit using complex parameters with the assumption $C_{9^{\prime}}^{\mu\mu}=-C_{10^{\prime}}^{\mu\mu}$. No overlap region indicates that $C_{9^{\prime}}^{\mu\mu}$ and $C_{10^{\prime}}^{\mu\mu}$ can not accommodate combined explanation for $R_K$ and $R_{K^*}$.}
  \label{contour-RK}
\end{figure}
According to the definition of $R_K$ and $R_{K^*}$, the anomalies of $R_{K^{(*)}}$ indicate new physics contribution to $C_{9^{(\prime)}}^{ee}$, $C_{10^{(\prime)}}^{ee}$, $C_{9^{(\prime)}}^{\mu\mu}$ and $C_{10^{(\prime)}}^{\mu\mu}$. The solution of $R_{K^*}$ is favored by new physics coupling to muon instead of electron, with the consideration from other observables fit~\cite{Capdevila:2017bsm, Hurth:2017hxg, Aebischer:2019mlg}. Therefore, we set the new physics contribution related to electron is negligible (i.e., $C_9^{ee}=C_{10}^{ee}\sim 0$), and the new physics contributions to $R_K$ and $R_{K^*}$ come from the $C_9^{\mu\mu}$ and $C_{10}^{\mu\mu}$ in our framework. We show the fit to $R_{K^{(*)}}$ using $C_{9^{(\prime)}, 10^{(\prime)}}$ at the scale $\mu = 1\ \text{TeV}$ in Fig.~\ref{contour-RK}. The upper left panel gives favored regions for $C_9^{\mu\mu}$ versus $C_{10}^{\mu\mu}$ as real parameters. We also consider the recent measurements for the ratio $R_{K^0_S}$ and $R_{K^{*+}}$ \cite{LHCb:2021lvy}. Note that we combine the constraint from $B_s \to \mu \mu$ in the fit. The branching ratio of $B_s \to \mu \mu$ is measured to be $\text{Br}(B_s \to \mu \mu)^{\text{exp}}=(2.93 \pm 0.35)\times 10^{-9}$ \cite{Geng:2021nhg}, which is the combined result based on measurements from ATLAS, CMS and LHCb \cite{ATLAS:2018cur, CMS:2019bbr, LHCb:2021awg, LHCb:2021vsc}, while the SM prediction value is $\text{Br}(B_s \to \mu \mu)^{\text{SM}}=(3.63\pm 0.13)\times 10^{-9}$~\cite{Beneke:2019slt}. Taking the relation $C_9^{\mu\mu} =-C_{10}^{\mu\mu  }$ given by our model, the best fit point of $R_{K^{(*)}}$ is found at $C_9^{\mu\mu} =-C_{10}^{\mu\mu}=-0.39$, while the best fit point for $R_{K^0_S}$ and $R_{K^{*+}}$ is found at $C_9^{\mu\mu} =-C_{10}^{\mu\mu}=-0.74$. Combining these four experimental ratios and the branching ratio of $B_s \to \mu \mu$, the best fit piont of $C_9^{\mu\mu} =-C_{10}^{\mu\mu}$ is $-0.45$. The upper right panel shows the favored region for the complex case with the assumption $C_9^{\mu\mu} =-C_{10}^{\mu\mu}$. The bottom panels present the fit to $R_{K}$ and $R_{K^{*}}$ using $C_{9^{\prime}, 10^{\prime}}^{\mu\mu}$ and we find no common solution. Our results are comparable with the global analysis performed in Ref.~\cite{Alok:2022pjb}, where some related differential branching ratios and angular observables are included. Relevant analyses are also found in Ref.~\cite{Becirevic:2015asa, Carvunis:2021jga,Altmannshofer:2021qrr}.

Leptoquark $S_1$ doesn't contribute to $b \to s \ell^+ \ell^{-}$ at tree-level but provides contribution by box-diagrams. However, $R_K$ and $R_{K^*}$ anomalies cannot be fully accommodated with leptoquark $S_1$ only~\cite{Becirevic:2016oho, Angelescu:2018tyl}. In our model, we expect that the contributions to solve $R_K$ and $R_{K^*}$ anomalies come dominantly from leptoquark $S_3$. The corresponding Wilson coefficients are given by
\begin{equation}
C_9^{\ell \ell}=-C_{10}^{\ell \ell}=\frac{\pi v^2}{V_{tb} V_{ts}^* \alpha_{\text{em}}}\frac{(V^Ty_{3L})^{3\ell}(V^{\dagger}y_{3L}^*)^{2\ell}}{m_{S_3}^2}\,.
\end{equation}
Leptoquark $\tilde{R}_2$ can also generate contribution to the process $b \to s \ell^+ \ell^{-}$ at tree-level by $C_{9^{\prime}}$ and $C_{10^{\prime}}$ terms. The corresponding Wilson coefficients of $\tilde{R}_2$ contribution are given by
\begin{equation}
C_{9^\prime}^{\ell \ell}=-C_{10^\prime}^{\ell \ell}=-\frac{\pi v^2}{2V_{tb} V_{ts}^* \alpha_{\text{em}}}\frac{y_{2L}^{2\ell}y_{2L}^{*3\ell}}{m_{R_2}^2}\,.
\end{equation}
It is noted that the parameter space to explain $R_K$ is incompatible with $R_{K^*}$ if one only use $C_{9^{\prime}}$ and $C_{10^{\prime}}$, as shown in the bottom panel of Fig.~\ref{contour-RK}.
\subsection{$R_D$ and $R_{D^*}$}
The next lepton flavor universality violation observables we consider are $R_D$ and $R_{D^*}$, which are induced by charged current transitions $b\to c \ell \bar{\nu}_{\ell}$ and defined as 
\begin{equation} 
R_{D}=\frac{\text{Br}(B \to D \tau \bar{\nu})}{\text{Br}(B \to D \ell \bar{\nu})}\,, \qquad
R_{D^{*}}=\frac{\text{Br}(B \to D^* \tau \bar{\nu})}{\text{Br}(B \to D^* \ell \bar{\nu})}\,,
\end{equation}
where $\ell$ denotes electron $e$ or muon $\mu$. The predicted values of these two observed quantities in the SM are~\cite{Bigi:2016mdz, Bernlochner:2017jka, Jaiswal:2017rve, Bigi:2017jbd}
\begin{equation}
R_D^{\rm{SM}} = 0.299 \pm 0.003\,, \qquad R_{D^*}^{\rm{SM}} = 0.258 \pm 0.003\,.
\end{equation}
These two observables have been measured independently by several collaborations, including Babar~\cite{BaBar:2012obs, BaBar:2013mob}, Belle~\cite{Belle:2015qfa, Belle:2016dyj} and LHCb~\cite{LHCb:2015gmp, LHCb:2017smo, LHCb:2017rln}. The average values by combining these measurements are given by~\cite{HFLAV:2019otj}
\begin{equation}
R_D^{\rm{exp}} = 0.340 \pm 0.027 \pm 0.013\,, \qquad R_{D^*}^{\rm{exp}} = 0.295 \pm 0.011 \pm 0.008\,,
\end{equation}
which exceed the SM predictions by 1.4$\sigma$ and 2.5$\sigma$ respectively. To confront the leptoquarks contributions with the above experimental data, we consider the following effective Hamiltonian,
\begin{equation}
\begin{split}
\mathcal{H}_{\text{eff}}=\frac{4 G_F}{\sqrt{2}} V_{cb}& \Big[g_{V_L} (\bar{c}_L \gamma_{\mu} b_L)(\bar{\tau}_L \gamma^{\mu} \nu_{L}) + g_{S_L} (\bar{c}_R b_L)(\bar{\tau}_R \nu_{L}) \\
&+ g_T (\bar{c}_R \sigma_{\mu\nu} b_L)(\bar{\tau}_R \sigma^{\mu\nu} \nu_{L}) \Big]+\text{h.c.}\,.
\end{split}
\end{equation}
\begin{figure}[!htb]
  \begin{minipage}{0.48\linewidth}
    \centerline{\includegraphics[width=1\textwidth]{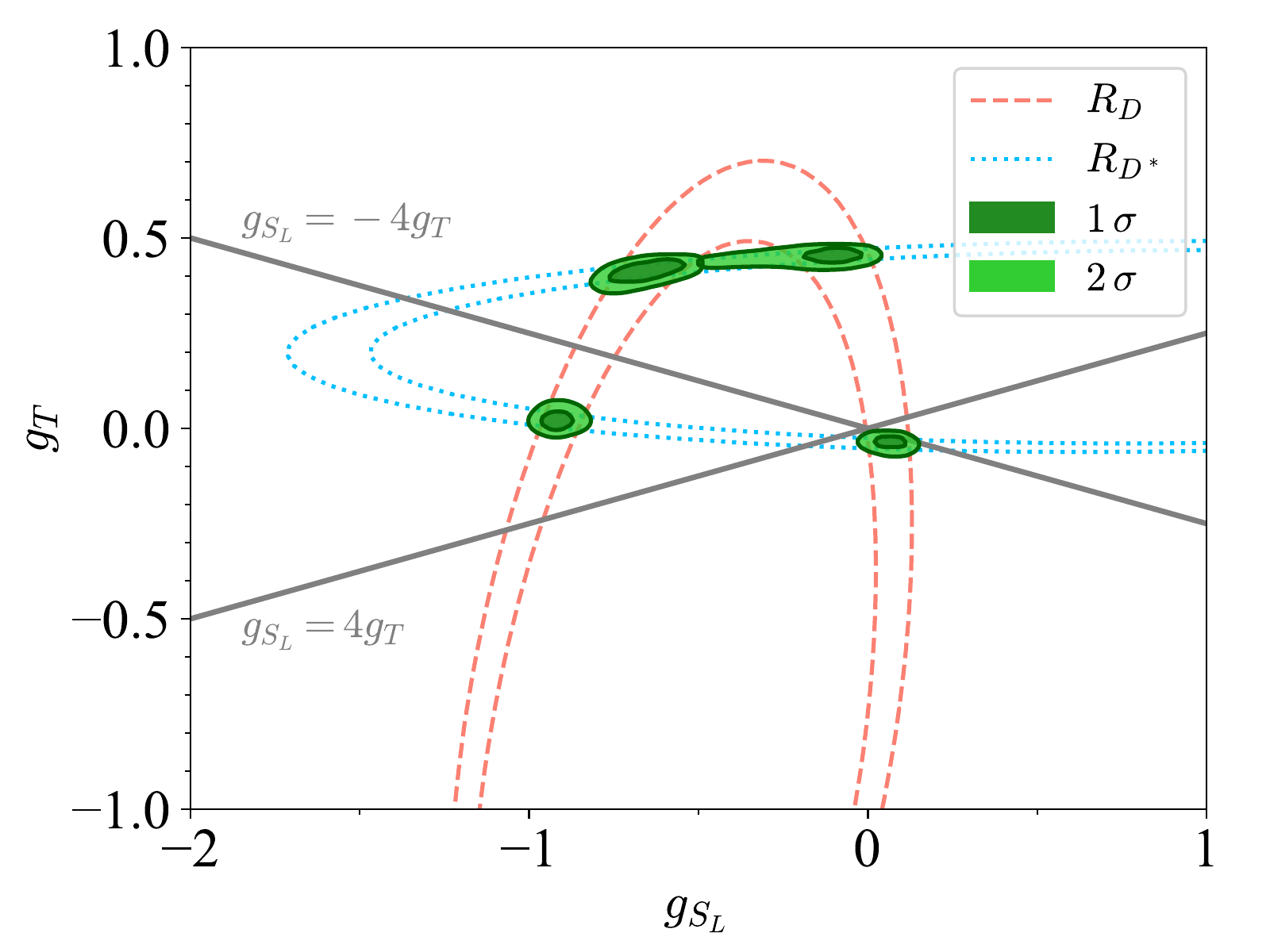}}
  \end{minipage}
  \begin{minipage}{0.48\linewidth}
    \centerline{\includegraphics[width=1\textwidth]{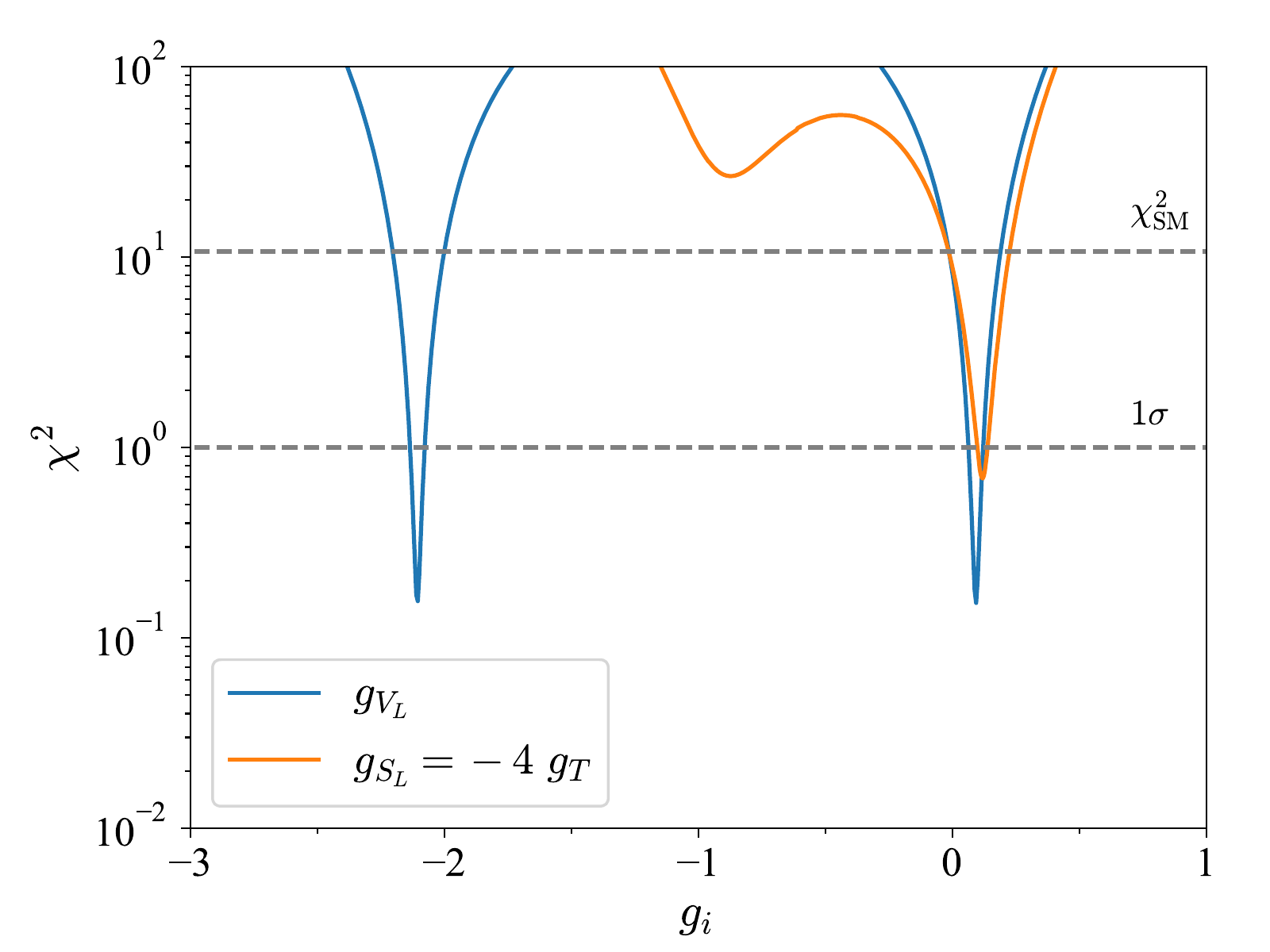}}
  \end{minipage}
  \begin{minipage}{0.48\linewidth}
    \centerline{\includegraphics[width=1\textwidth]{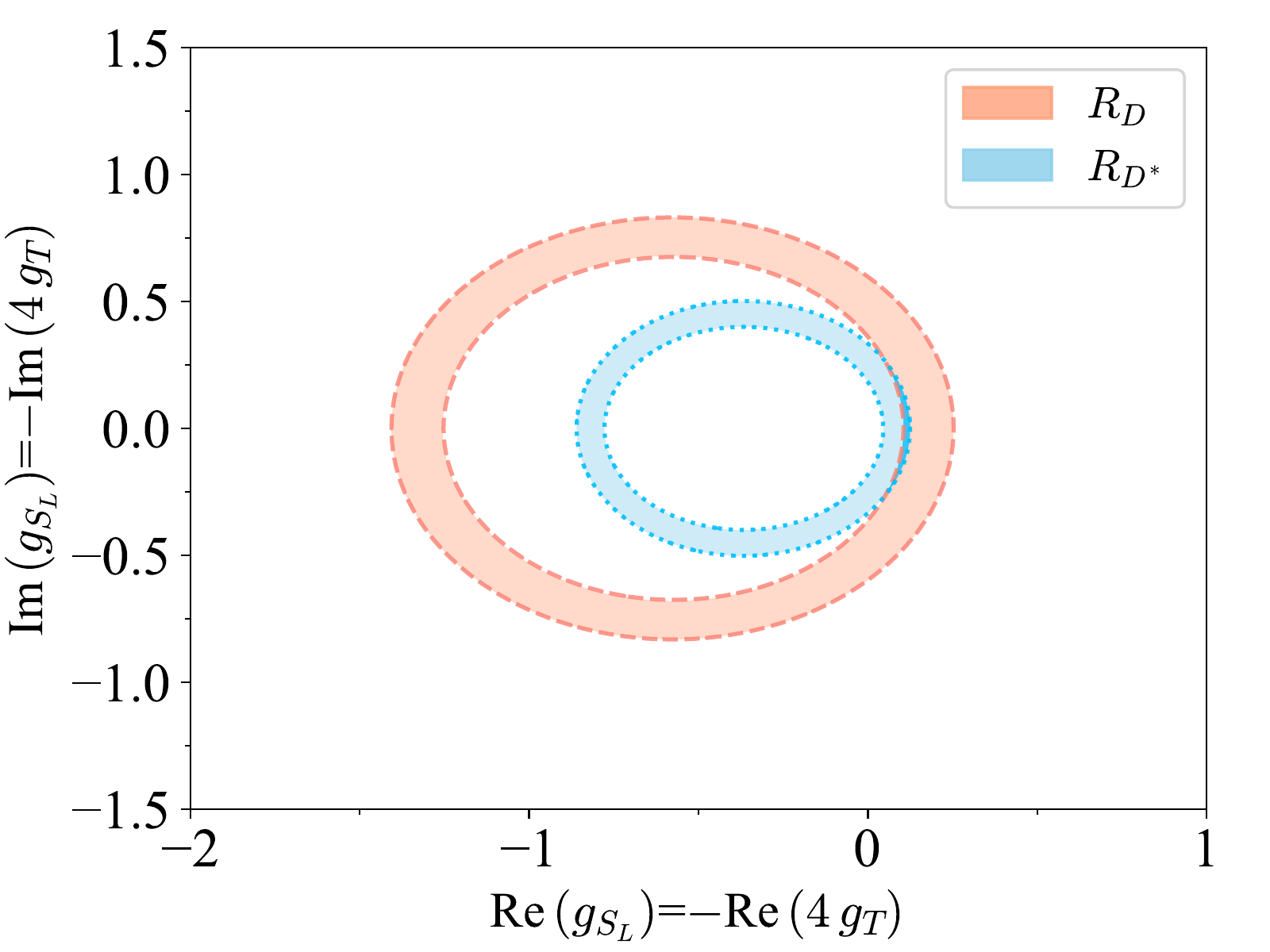}}
  \end{minipage}
  \begin{minipage}{0.48\linewidth}
    \centerline{\includegraphics[width=1\textwidth]{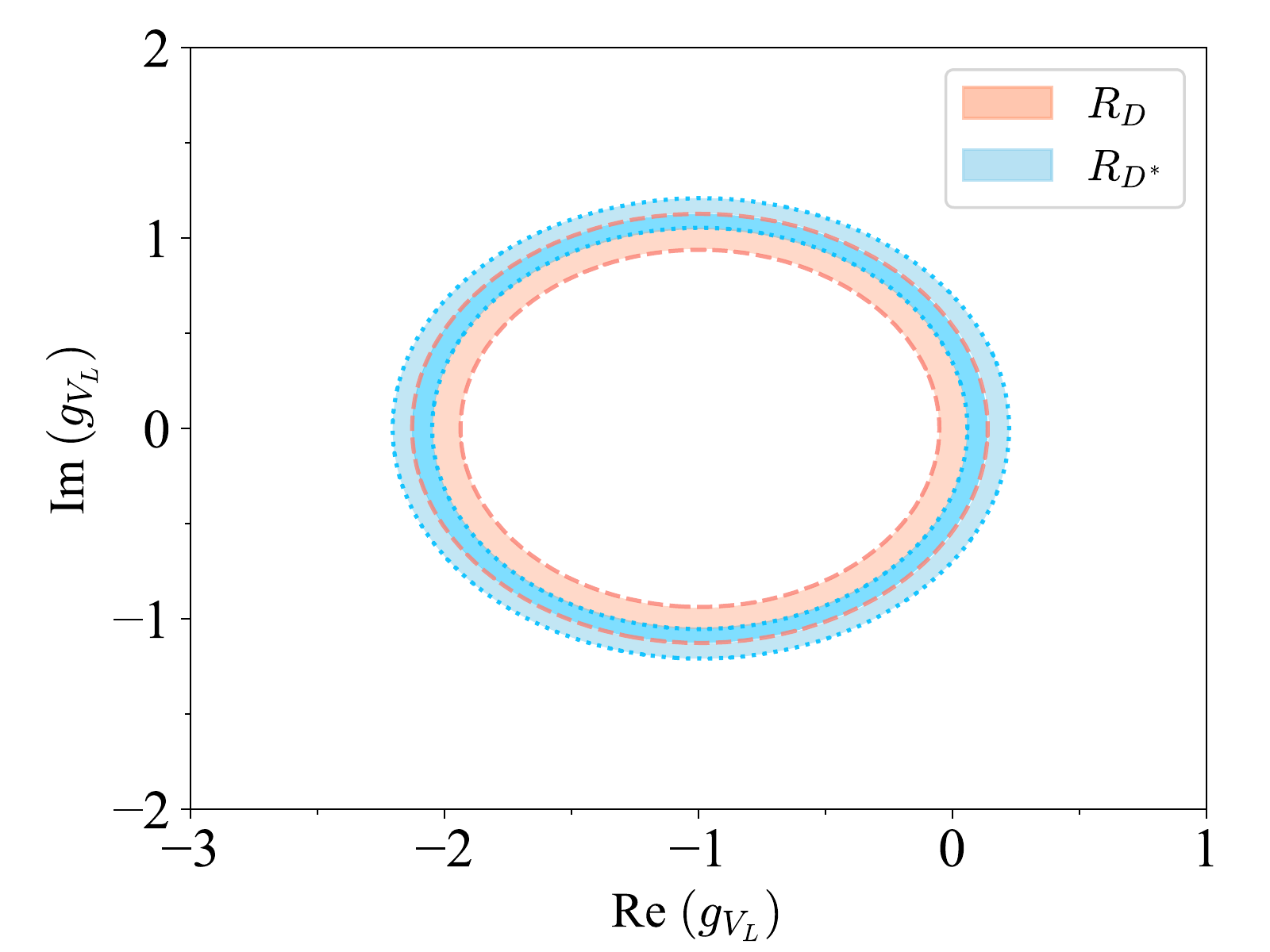}}
  \end{minipage}
  \caption{Upper left: Contour plot of the fit for $R_D$ and $R_{D^*}$ in the plane of $g_{S_L}$ versus $g_T$ at the scale $\mu=1\,\text{TeV}$. The dashed orange and dotted blue contours represent the $1\sigma$ allowed regions that explain $R_D$ and $R_{D^*}$, respectively. The dark (light) green region corresponds to the $1\sigma$ ($2\sigma$) allowed region that explains both simultaneously. Upper right: The $\chi^2$ values to fit both $R_D$ and $R_{D^*}$ when using $g_{V_L}$ and $g_{S_L}=-4g_T$ respectively. Bottom: Two plots correspond to the complex planes of $g_{S_L}=-4g_T$ and $g_{V_L}$, respectively. The orange and blue regions represent the $1\sigma$ allowed region of $R_D$ and $R_{D^*}$, respectively. The deeper blue region corresponds to the overlap scenario where accommodates combined explanation for $R_D$ and $R_{D^*}$.}
  \label{contour-RD}
\end{figure}
\noindent In the Fig.~\ref{contour-RD}, we show the fit of $g_{S_L}, g_T$ and $g_{V_L}$ favored region to explain $R_D$ and $R_{D^*}$ anomalies at the scale of $1$ TeV. The upper left panel presents the fit using real parameters $g_{S_{L}}$ and $g_{T}$. With the relation of $g_{S_L}=-4g_{T}$, which is in our model, the best fit point is $g_{S_L}=-4g_T = 0.12$ and the allowed $1\,\sigma$ range is $g_{S_L}=-4g_T \in [0.08, 0.16]$. If we solely consider the Wilson coefficient $g_{V_L}$, the best fit point is $g_{V_L} = 0.08(-2.07)$ and the allowed $1\, \sigma$ range is $g_{V_L} \in [0.07, 0.10] \cup [-2.10, -2.05]$. We show the $\chi^{2}$ values to fit both $R_{D}$ and $R_{D^{*}}$ in the upper right panel. We also present the fit result when the coefficients are taken as complex numbers. Comprehensive analyses including the ratio $R_{J/\psi}$, the longitudinal polarization of the $P_{\tau}(D^*)$ and $F_{L}^{D^*}$ can be found in Ref.~\cite{Murgui:2019czp, Shi:2019gxi, Cheung:2020sbq}. The best fit values in this work agree with theirs in the $1\,\sigma$ allowed range.

In the model, both $S_1$ and $S_3$ give contributions to $b \to c \tau \bar{\nu}$ at tree-level, while $\tilde{R}_2$ does not. After Fierz transformation to relevant effective Lagrangian, $g_{S_L}$ and $g_T$ have relation $g_{S_L}= -4g_T$. The corresponding Wilson coefficient of $S_1$ contributions are given by
\begin{align}
&g_{V_L}^{\ell} = \frac{v^2}{4V_{cb}}\frac{(V^Ty_{1L})^{3\ell}y_{1L}^{*23}}{m_{S_1}^2}\,, \\
&g_{S_L}^{\ell} = -4g_T^{\ell}=-\frac{v^2}{4V_{cb}}\frac{(V^Ty_{1L})^{3\ell}y_{1R}^{*23}}{m_{S_1}^2}\,. 
\end{align}
The contribution from leptoquark $S_3$ gives
\begin{equation}
g_{V_L}^{\ell} = -\frac{v^2}{4V_{cb}}\frac{(V^Ty_{3L})^{3\ell}y_{3L}^{*23}}{m_{S_3}^2}\,.
\end{equation}
However the contributions of $g_{V_L}^\ell$ from both leptoquarks $S_1$ and $S_3$ can not explain the anomalies of $R_{D^{(*)}}$ since its favored parameters space is incompatible with $B$ meson decay process $B \to K \nu \bar{\nu}$. To explain the anomalies of $R_{D^{(*)}}$, it is required that $y_{1L,3L}^{33}y_{1L,3L}^{23}\sim 0.1$, but the products of couplings are strongly constrained by the process $B \to K \nu \bar{\nu}$ with $|y_{1L,3L}^{33}y_{1L,3L}^{23}| \lesssim 0.03$. Therefore we focus on the Wilson coefficients $g_{S_L}^\ell$ and $g_T^\ell$ contribution from the Leptoquark $S_1$ to explain the anomalies of $R_{D^{(*)}}$.

\subsection{The anomalous magnetic moments of charged leptons}
The last observable anomalies we consider are the anomalous magnetic moments of charged leptons, including electron and muon, which both exist long-standing discrepancy between the SM predictions and experimental measurements. The recent combined result of Fermilab~\cite{Abi:2021gix} and BNL~\cite{Bennett:2006fi} increases the tension of muon $(g-2)$, which gives a $4.2\, \sigma$ level deviation from the SM prediction. The precise discrepancy between the SM predictions and experimental values reads~\cite{Aoyama:2017uqe, Keshavarzi:2018mgv, Davier:2019can, Aoyama:2020ynm,Aoyama:2012wk,Aoyama:2019ryr,Czarnecki:2002nt,Gnendiger:2013pva,Davier:2017zfy,Colangelo:2018mtw,Hoferichter:2019mqg,Keshavarzi:2019abf,Kurz:2014wya,Melnikov:2003xd,Masjuan:2017tvw,Colangelo:2017fiz,Hoferichter:2018kwz,Gerardin:2019vio,Bijnens:2019ghy,Colangelo:2019uex,Blum:2019ugy,Colangelo:2014qya,Abi:2021gix}
\begin{align}
&\Delta a_e = a_e^{\text{exp}} - a_e^{\text{SM}} = -(8.7 \pm 3.6) \times 10^{-13}\,, \\
&\Delta a_{\mu} = a_{\mu}^{\text{exp}} - a_{\mu}^{\text{SM}} = (2.51 \pm 0.59) \times 10^{-9}\,.
\end{align}

\begin{figure}[!t]
\centering
\begin{tikzpicture}[line width=1.2 pt, scale=1.5, >=latex]
	\draw[fermion] (-1,0)--(0,0);
	\draw[fermion] (0,0)--(2,0);
	\draw[scalarnoarrow] (0,0) arc (180:90:1);
	\draw[scalarnoarrow] (2,0) arc (0:90:1);
	\draw[vector] (1.5,1.2)--(2.5,1.8);
	\draw[fermion] (2,0)--(3,0);
	\node at (-1,-0.3) {$\ell$};
	\node at (1,-0.3) {$q\,(q^C)$};
	\node at (3,-0.3) {$\ell^{(\prime)}$};
	\node at (0,1) {$S\, (S^*)$};
	\node at (2.7,1.8) {$\gamma$};
\end{tikzpicture}
\caption{One-loop diagram contributing to the charged leptons anomalous magnetic moments and the flavor changing process $\ell\to\ell^{\prime}\gamma$.}
\label{g-2}
\end{figure}
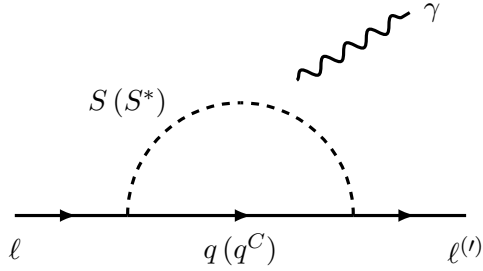

We start to discuss the contributions to $(g-2)_{\ell}$ from general scalar Leptoquark interactions, which is described by~\cite{Cheung:2001ip, Dorsner:2017wwn}
\begin{align}
\mathcal{L}^{F=0}&=\overline{q}_i(y_R^{ij} P_R + y_L^{ij} P_L)\ell_j S + \text{h.c.}\,, \\
\mathcal{L}^{|F|=2}&=\overline{q^C_i}(y_R^{\prime ij} P_R + y_L^{\prime ij} P_L)\ell_j S + \text{h.c.}\,.
\end{align}
Here $q^i$ denotes quark, $\ell$ denotes charged leptons, $S$ stands for leptoquarks and $F$ is the fermion number. The contributions to $\Delta a_{\ell}\equiv(g-2)_{\ell}/2$ from $F=0$ terms are illustrated in Fig.~\ref{g-2} and given by
\begin{equation}
\Delta a_{\ell}=-\frac{3m_{\ell}}{8 \pi^2 m_S^2} \sum_{q} \left[m_{\ell}(|y_R^{q\ell}|^2 + |y_L^{q\ell}|^2)F(x) + m_q \text{Re}(y_L^{*q\ell} y_R^{q\ell})G(x) \right]\,,
\label{g-2l}
\end{equation}
where
\begin{equation}
\begin{split}
F(x) &= Q_S\, f_S(x) - f_F(x)\,, \\
G(x) &= Q_S\, g_S(x) - g_F(x)\,,
\end{split}
\label{g-2FF}
\end{equation}
and the loop functions are calculated by following formulas,
\begin{align}
&f_S(x)=\frac{x+1}{4(1-x)^2} + \frac{x \ln x}{2(1-x)^3}\,,   \nonumber \\
&f_F(x)=\frac{2+5x-x^2}{12(1-x)^3} + \frac{x \ln x}{2(1-x)^4}\,, \nonumber  \\
&g_S(x)=\frac{-1}{1-x} - \frac{\ln x}{(1-x)^2} \,, \nonumber \\
&g_F(x)=\frac{x-3}{2(1-x)^2} - \frac{\ln x}{(1-x)^3} \,,\label{loopfun}
\end{align}
where $x=m_q^2/m_S^2$ and $Q_{S}$ is the charge of leptoquark $S$. The $|F|=2$ scalar leptoquarks contribution can be obtained by changing the couplings $y\to y^{\prime}$ in Eq.~(\ref{g-2l}).  It is noted that the no-chiral scalar leptoquarks which have both left-handed and right-handed couplings to quarks can give a chiral-enhanced contributions to $\Delta a_{\ell}$ by the quark masses. This is revealed by Eq.~(\ref{g-2l}), in which the first term is proportional to the lepton mass, while the second term is proportional to the internal quark mass. Besides, it is worthy to notice that only the second term in Eq.~(\ref{g-2l}) can provide different sign contribution, since the deviation $\Delta a_{e}$ and $\Delta a_{\mu}$ have opposite sign. Thereby among all the scalar leptoquarks, only singlet $S_1$ or doublet $R_2$ could provide solution to explain $\Delta a_e$ and $\Delta a_{\mu}$ simultaneously. However, the constraint from the branching ratio of $\mu \to e \gamma$ excludes the one internal quark, such as the top quark, dominating solution~\cite{Crivellin:2018qmi}.  In our model, we choose the scenario that the contributions to  $\Delta a_{e}$ and $\Delta a_{\mu}$ come from different quarks. The new contribution is mainly coming from the leptoquark $S_1$ mediated loop and the contribution to $\Delta a_\ell$ is given by
\begin{equation}
\Delta a_{\ell} \simeq -\sum_q \frac{3m_{\ell} m_q}{8 \pi^2 m_{S_1}^2}  \text{Re}\Big(y_{1R}^{*q\ell} y_{1L}^{q\ell}\Big)\bigg[\frac{7}{6}+\frac{2}{3}\ln x \bigg]\,.
\end{equation}

\section{Low energy constraints}
\label{constraints}
In the previous section, we have discussed the solution to the $B$-physics anomalies, $R_{K^{(*)}}$ and $R_{D^{(*)}}$, and the anomalous charged lepton magnetic moments, $\Delta a_e$ and $\Delta a_{\mu}$. The model also gives rise to various flavor violating processes and rare meson decays, which are severely constrained by current experiments. In this section, we summarize the most stringent low-energy processes and give the relevant constraints to the leptoquark couplings in the model.

\subsection{$\ell \to \ell^{\prime} \gamma$ processes}
The lepton flavor violation $\ell \to \ell^{\prime} \gamma$ processes, such as $\mu \to e \gamma$, $\tau \to e \gamma$ and $\tau \to \mu \gamma$, can be induced via the one-loop diagrams shown in Fig.~\ref{g-2}. The non-chiral leptoquark $S_1$ contribution to $\ell \to \ell^{\prime} \gamma$ processes is enhanced by the quark mass. On the contrary, the chiral leptoquarks $\tilde{R}_2$ and $S_3$ induce $\ell \to \ell^{\prime} \gamma$ processes without chiral enhancement. The current experimental limits on the lepton flavor violation $\ell \to \ell^{\prime} \gamma$ processes are summarized as following~\cite{TheMEG:2016wtm, Aubert:2009ag},
\begin{align}
&\text{Br}(\mu \to e \gamma)<4.2 \times 10^{-13}\,,  \\
&\text{Br}(\tau \to e \gamma)<3.3 \times 10^{-8}\,, \\
&\text{Br}(\tau \to \mu \gamma)<4.4 \times 10^{-8}\,.
\end{align}
The branching ratio of the process $\ell \to \ell^{\prime} \gamma$ mediated by the leptoquarks can be calculated by the following formula,
\begin{equation}
\text{Br}(\ell \to \ell^{\prime} \gamma)=\frac{\alpha_{\text{em}}(m_{\ell}^2-m_{\ell^{\prime}}^2)^3}{4m_{\ell}^3\,\Gamma(\ell)}\Big(|\sigma_R^{\ell\ell^{\prime}}|^2 + |\sigma_L^{\ell\ell^{\prime}}|^2 \Big)\,,
\end{equation}
where $\Gamma(\ell)$ is the total decay width of the lepton $\ell$ and the form factors $\sigma_L^{\ell\ell^{\prime}}$ and $\sigma_R^{\ell\ell^{\prime}}$ originating from $S_1$, $\tilde{R}_2$ and $S_3$ contribution are calculated as
\begin{align}
\sigma_{L,S_1}^{\ell\ell^{\prime}}=&\frac{3}{16\pi^2m_{S_1}^2}\sum_{q=u,c,t}  \bigg\{\Big[m_{\ell}\,y_{1R}^{q\ell} y_{1R}^{*q\ell^{\prime}}+ m_{\ell^{\prime}}\,y_{1L}^{q\ell} y_{1L}^{*q\ell^{\prime}}\Big]\big[\frac{1}{3}f_S(x)-f_F(x)\big] \nonumber \\
& + m_q\big(y_{1L}^{q\ell} y_{1R}^{*q\ell^{\prime}}\big)\big[\frac{1}{3}\textsl{g}_S(x)-\textsl{g}_F(x) \big] \bigg\}\,, \label{ltolg1} \\
\sigma_{R,S_1}^{\ell\ell^{\prime}}=&\frac{3}{16\pi^2m_{S_1}^2}\sum_{q=u,c,t}  \bigg\{\Big[m_{\ell}\,y_{1L}^{q\ell} y_{1L}^{*q\ell^{\prime}}+ m_{\ell^{\prime}}\,y_{1R}^{q\ell} y_{1R}^{*q\ell^{\prime}}\Big]\big[\frac{1}{3}f_S(x)-f_F(x)\big] \nonumber \\
& + m_q\big(y_{1R}^{q\ell} y_{1L}^{*q\ell^{\prime}}\big)\big[\frac{1}{3}\textsl{g}_S(x)-\textsl{g}_F(x) \big] \bigg\}\,, \label{ltolg2}\\
\sigma_{L,\tilde{R}_2}^{\ell\ell^{\prime}}=&\frac{3}{16\pi^2m_{R_2}^2} \sum_{q=d,s,b}m_{\ell^{\prime}}\,y_{2L}^{q\ell} y_{2L}^{*q\ell^{\prime}}\big[\frac{2}{3}f_S(x)-f_F(x)\big]\,, \label{ltolg3} \\
\sigma_{R,\tilde{R}_2}^{\ell\ell^{\prime}}=&\frac{3}{16\pi^2m_{R_2}^2} \sum_{q=d,s,b} m_{\ell}\,y_{2L}^{q\ell} y_{2L}^{*q\ell^{\prime}}\big[\frac{2}{3}f_S(x)-f_F(x)\big]\,,\label{ltolg4}\\
\sigma_{L,S_3}^{\ell\ell^{\prime}}=&\frac{3}{16\pi^2m_{S_3}^2}  \bigg\{\sum_{q=d,s,b}2m_{\ell^{\prime}}\,(V^Ty_{3L})^{q\ell} (V^{\dagger}y_{3L}^*)^{q\ell^{\prime}}\big[\frac{4}{3}f_S(x)-f_F(x)\big]\nonumber \\
&+\sum_{q=u,c,t}m_{\ell^{\prime}}\,y_{3L}^{q\ell} y_{3L}^{*q\ell^{\prime}}\big[\frac{1}{3}f_S(x)-f_F(x)\big]  \bigg\}\,, \\
\sigma_{R,S_3}^{\ell\ell^{\prime}}=&\frac{3}{16\pi^2m_{S_3}^2}  \bigg\{\sum_{q=d,s,b}2m_{\ell}\,(V^Ty_{3L})^{q\ell} (V^{\dagger}y_{3L}^*)^{q\ell^{\prime}}\big[\frac{4}{3}f_S(x)-f_F(x)\big]  \nonumber \\
& +\sum_{q=u,c,t}m_{\ell}\,y_{3L}^{q\ell} y_{3L}^{*q\ell^{\prime}}\big[\frac{1}{3}f_S(x)-f_F(x)\big]\bigg\}\,,
\end{align}
where the loop functions $f_{S,F}(x)$ and $g_{S,F}(x)$ are defined in Eqs.~(\ref{loopfun}). The terms above proportional to $m_q$ arising from the non-chiral leptoquark $S_1$ give an enhancement and the corresponding couplings are more severely limited. Whereas the chiral leptoquarks $\tilde{R}_2$ and $S_3$ only consist of the terms proportional to $m_{\ell^{(\prime)}}$ and get weaker limits. To get the constraints on the couplings, we assume that only the relevant term dominates the contribution. The relevant constraints on the leptoquark Yukawa couplings are summarized in Table~{\ref{ltolgconstrain}}.
\begin{table}[ht]
\centering
 \begin{tabular}{|c|l|}\hline
 Process               & \multicolumn{1}{c|}{Constraints}  \\ \hline
 \multirow{5}{*}{$\mu \to e \gamma$}    & $|y_{1L}^{12}y_{1R}^{*11}|, |y_{1L}^{*11}y_{1R}^{12}| < 3.57 \times 10^{-4} \left(\frac{m_{S_1}}{\text{TeV}}\right)^2$ \\ 
  & $|y_{1L}^{22}y_{1R}^{*21}|, |y_{1L}^{*21}y_{1R}^{22}| < 1.29 \times 10^{-6} \left(\frac{m_{S_1}}{\text{TeV}}\right)^2$   \\ 
  & $|y_{1L}^{32}y_{1R}^{*31}|, |y_{1L}^{*31}y_{1R}^{32}| < 5.38 \times 10^{-8} \left(\frac{m_{S_1}}{\text{TeV}}\right)^2$  \\ 
  & $|y_{1L}^{i2}y_{1L}^{*i1}|,|y_{1R}^{i2}y_{1R}^{*i1}|,|y_{3L}^{i2}y_{3L}^{*i1}| < 1.31 \times 10^{-3} \left(\frac{m_{S_1/S_3}}{\text{TeV}}\right)^2$  \\ 
  & $|(V^Ty_{3L})^{i2} (V^{\dagger}y_{3L}^*)^{i1}| < 3.98 \times 10^{-4} \left(\frac{m_{S_3}}{\text{TeV}}\right)^2$ \\ \hline 
  \multirow{5}{*}{$\tau \to e \gamma$}    & $|y_{1L}^{13}y_{1R}^{*11}|, |y_{1L}^{*11}y_{1R}^{13}| < 3.99 \left(\frac{m_{S_1}}{\text{TeV}}\right)^2$ \\ 
  & $|y_{1L}^{23}y_{1R}^{*21}|, |y_{1L}^{*21}y_{1R}^{23}| < 1.45\times 10^{-2} \left(\frac{m_{S_1}}{\text{TeV}}\right)^2$   \\ 
  & $|y_{1L}^{33}y_{1R}^{*31}|, |y_{1L}^{*31}y_{1R}^{33}| < 6.02 \times 10^{-4} \left(\frac{m_{S_1}}{\text{TeV}}\right)^2$  \\ 
  & $|y_{1L}^{i3}y_{1L}^{*i1}|,|y_{1R}^{i3}y_{1R}^{*i1}|,|y_{3L}^{i3}y_{3L}^{*i1}| < 0.874 \left(\frac{m_{S_1/S_3}}{\text{TeV}}\right)^2$ \\ 
  & $|(V^Ty_{3L})^{i3} (V^{\dagger}y_{3L}^*)^{i1}| < 0.240 \left(\frac{m_{S_3}}{\text{TeV}}\right)^2$ \\ \hline 
  \multirow{5}{*}{$\tau \to \mu \gamma$}  & $|y_{1L}^{13}y_{1R}^{*12}|, |y_{1L}^{*12}y_{1R}^{13}| < 4.61   \left(\frac{m_{S_1}}{\text{TeV}}\right)^2$ \\ 
  & $|y_{1L}^{23}y_{1R}^{*22}|, |y_{1L}^{*22}y_{1R}^{23}| < 1.67 \times 10^{-2} \left(\frac{m_{S_1}}{\text{TeV}}\right)^2$  \\ 
  & $|y_{1L}^{33}y_{1R}^{*32}|, |y_{1L}^{*32}y_{1R}^{33}|| < 6.95 \times 10^{-4} \left(\frac{m_{S_1}}{\text{TeV}}\right)^2$  \\ 
  & $|y_{1L}^{i3}y_{1L}^{*i2}|,|y_{1R}^{i3}y_{1R}^{*i2}|,|y_{3L}^{i3}y_{3L}^{*i2}| < 1.01 \left(\frac{m_{S_1/S_3}}{\text{TeV}}\right)^2$ \\
  & $|(V^Ty_{3L})^{i3} (V^{\dagger}y_{3L}^*)^{i2}| < 0.278 \left(\frac{m_{S_3}}{\text{TeV}}\right)^2$ \\ \hline
  \multirow{5}{*}{$\mu \text{Au} \to e \text{Au}$} & $|y_{1L}^{12} y_{1L}^{*11}|,|y_{1R}^{12} y_{1R}^{*11}| < 4.20\times 10^{-6} \left(\frac{m_{S_1}}{\text{TeV}}\right)^2$ \\ 
  & $|y_{1L}^{12} y_{1R}^{*11}|,|y_{1R}^{12} y_{1L}^{*11}| < 8.12 \times 10^{-6} \left(\frac{m_{S_1}}{\text{TeV}}\right)^2$ \\ 
  & $|y_{2L}^{12}y_{2L}^{*11}| < 3.40 \times 10^{-6} \left(\frac{m_{R_2}}{\text{TeV}}\right)^2$ \\
  & $|y_{3L}^{12}y_{3L}^{*11}| < 2.14 \times 10^{-6} \left(\frac{m_{S_3}}{\text{TeV}}\right)^2$ \\
  & $|(V^Ty_{3L})^{12} (V^{\dagger}y_{3L}^*)^{11}| < 1.70 \times 10^{-6} \left(\frac{m_{S_3}}{\text{TeV}}\right)^2$ \\ \hline
 \end{tabular}
  \caption{Upper limits on the leptoquark couplings from the processes $\ell \to \ell^{\prime} \gamma$ and $\mu\text{Au}\to e \text{Au}$.}
 \label{ltolgconstrain}
\end{table}

\subsection{$\mu - e$ conversion in nuclei}
Besides the charged lepton flavor violating radiative decay processes, $\mu - e$ conversion in nuclei is also a rare process providing stringent constraints on the strength of leptoquark interactions. The current experimental search on $\mu - e$ conversion using gold nucleus provides the most stringent upper limits and the upper bound to the branching ratio is set by the SINDRUM experiment as~\cite{SINDRUMII:2006dvw}
\begin{equation}
\text{Br}(\mu - e)_{\text{Au}}=\frac{\Gamma(\mu - e)_{\text{Au}}}{\Gamma_{\text{capture}}} <7 \times 10^{-13}\,,
\end{equation}
where the $\Gamma_{\text{capture}}=8.6 \times 10^{-18}$ GeV denotes the muon capture rate by gold nucleus~\cite{Suzuki:1987jf}. The $\mu - e$ conversion rate in nuclei can be calculated by following formula~\cite{Kitano:2002mt, Arganda:2007jw}
\begin{equation}
\Gamma(\mu - e) = 2 G_F^2 m_{\mu}^{5}\left| \tilde{g}_{LS}^{(p)}S^{(p)}+\tilde{g}_{LS}^{(n)}S^{(n)}+\tilde{g}_{LV}^{(p)}V^{(p)} + \tilde{g}_{LV}^{(n)}V^{(n)} \right|^2 +(L\to R)\,.
\end{equation}
The overlap integral values of gold nucleus are $S^{(p)}=0.0523$, $S^{(n)}=0.0610$, $V^{(p)}=0.0859$, $V^{(n)}=0.108$~\cite{Kitano:2002mt}. With the effective Lagrangian given in Eq.~(\ref{EFL1}), the coupling constants $\tilde{g}$ are defined as
\begin{align}
&\tilde{g}_{LS,RS}^{(p)}=\sum_q G_{S}^{(q,p)}\frac{1}{2} (g_{S,q}^{LL,RR})^{ii,12}\,, \\
&\tilde{g}_{LS,RS}^{(n)}=\sum_q G_{S}^{(q,n)}\frac{1}{2} (g_{S,q}^{LL,RR})^{ii,12}\,, \\
&\tilde{g}_{LV}^{(p)}=\left[ (g_{V,u}^{LL})^{11,12} + (g_{V,u}^{RL})^{11,12}\right] + \frac{1}{2} \left[ (g_{V,d}^{LL})^{11,12} + (g_{V,d}^{RL})^{11,12}\right]\,, \\
&\tilde{g}_{RV}^{(p)}=\left[ (g_{V,u}^{RR})^{11,12} + (g_{V,u}^{LR})^{11,12}\right] + \frac{1}{2} \left[ (g_{V,d}^{RR})^{11,12} + (g_{V,d}^{LR})^{11,12}\right]\,, \\
&\tilde{g}_{LV}^{(n)}=\frac{1}{2}\left[ (g_{V,u}^{LL})^{11,12} + (g_{V,u}^{RL})^{11,12}\right] + \left[ (g_{V,d}^{LL})^{11,12} + (g_{V,d}^{RL})^{11,12}\right]\,, \\
&\tilde{g}_{RV}^{(n)}=\frac{1}{2}\left[ (g_{V,u}^{RR})^{11,12} + (g_{V,u}^{LR})^{11,12}\right] + \left[ (g_{V,d}^{RR})^{11,12} + (g_{V,d}^{LR})^{11,12}\right]\,,
\end{align}
where the coefficients of scalar operators are $G_S^{u,p}=G_S^{d,n}=5.1$, $G_S^{d,p}=G_S^{u,n}=4.3$ and $G_S^{s,p}=G_S^{s,n}=2.5$~\cite{Kosmas:2001mv}. The bounds on the leptoquark couplings from $\text{Br}(\mu-e)_{\text{Au}}$ are summarized in Table~\ref{ltolgconstrain}.

\subsection{Rare meson leptonic decays}
Introducing leptoquarks could induce meson rare decay processes. In this subsection, we consider the relevant $B_s$ meson rare leptonic decays that include leptonic conserving decays, $B_{s} \to \mu^+ \mu^-/\tau^+ \tau^-$, and leptonic flavor violation decay $B_{s} \to \mu^\pm \tau^\mp$. The corresponding 4-fermion operators $\mathcal{O}_{9^{(\prime)}}, \mathcal{O}_{10^{(\prime)}}$ are given in the Eq.~(\ref{Wilson910}). The recent experimental measurements of these processes are given by~\cite{Geng:2021nhg, LHCb:2017myy, LHCb:2019ujz}
\begin{align}
&\text{Br}(B_s \to \mu^{+} \mu^{-}) =(2.93 \pm 0.35)\times 10^{-9}\,, \\
&\text{Br}(B_s \to \tau^{+} \tau^{-}) <6.8\times 10^{-3}\,, \\
&\text{Br}(B_s \to \mu^{\pm} \tau^{\mp}) <1.4\times 10^{-5}\,.
\end{align}
Among these processes, only the $B_s \to \mu^+ \mu^-$ has been observed by the current experiments and the branching ratio agrees with SM prediction value at a level of $4\sigma$, $\text{Br}(B_s \to \mu \mu)^{\text{SM}}=(3.63\pm 0.13)\times 10^{-9}$~\cite{Beneke:2019slt}, while the current experiments only give upper bounds for the other two processes. The contribution to the decay width of a neutral meson to two charged leptons $P \to \ell^+ \ell^{\prime -}$ can be written as~\cite{Becirevic:2016zri}
\begin{align}
\Gamma_{P \to \ell^+ \ell^{\prime -}} = &\frac{1}{64 \pi^3} \frac{G_F^2 \alpha^2_{\text{em}}}{m_P^3} f_P^2 |V_{qj} V^*_{qi}|^2 \lambda_1^{1/2} \lambda_2^{1/2}  \nonumber \\ 
&\times \Bigg\{\lambda_1 \cdot \left|(m_{\ell}-m_{\ell^{\prime}}) \left(C_9^{ij\ell\ell^{\prime}} - C_{9^{\prime}}^{ij\ell\ell^{\prime}}\right)+\frac{m_P^2}{m_q+m_{q^{\prime}}}(C_S-C_S^{\prime}) \right|^2  \nonumber \\ 
&+ \lambda_2 \cdot \left|(m_{\ell}+m_{\ell^{\prime}}) \left(C_{10}^{ij\ell\ell^{\prime}} - C_{10^{\prime}}^{ij\ell\ell^{\prime}}\right) + \frac{m_P^2}{m_q+m_{q^{\prime}}}(C_P-C_P^{\prime}) \right|^2 \Bigg\}\,,
\end{align}
where $f_P$ is the meson decay constant, $\lambda_{1,2}=m_P^2-(m_{\ell} \pm m_{\ell^{\prime}})^2$ and $m_q, m_{q^{\prime}}$ are the masses of the valence quarks in the pseudoscalar meson $P$. It is noted that the lepton flavor conserving decay process $P \to \ell^+ \ell^-$ is independent of Wilson coefficients $C_{9^{(\prime)}}$.

\begin{table}[ht]
\centering
 \begin{tabular}{|c|l|}\hline
 Processes               & \multicolumn{1}{c|}{Constraints on the couplings} \\ \hline
  \multirow{2}{*}{$B^0_s \to \mu^+ \mu^-$}  & $y_{2L}^{32} y_{2L}^{*22} \in [-1.1,1.1] \times 10^{-4} \times \left( \frac{m_{R_2}}{\text{TeV}}\right)^2$ \\ 
   & $(V^Ty_{3L})^{22} (V^{\dagger}y_{3L}^*)^{32} \in [-5.8,5.3] \times 10^{-4} \times \left( \frac{m_{S_3}}{\text{TeV}}\right)^2$ \\ \hline 
 \multirow{2}{*}{$B^0_s \to \tau^+ \tau^-$}  & $y_{2L}^{33} y_{2L}^{*23} \in [-1.3,1.3] \times \left( \frac{m_{R_2}}{\text{TeV}}\right)^2$ \\ 
   & $(V^Ty_{3L})^{23} (V^{\dagger}y_{3L}^*)^{33}  \in [-0.63,0.63] \times \left( \frac{m_{S_3}}{\text{TeV}}\right)^2$ \\ \hline
 \multirow{2}{*}{$B^0_s \to \mu^{\pm} \tau^{\mp}$}  &  $y_{2L}^{32} y_{2L}^{*23}, y_{2L}^{33} y_{2L}^{*22} \in [-0.080,0.080] \times \left( \frac{m_{R_2}}{\text{TeV}}\right)^2$ \\ 
   & $(V^Ty_{3L})^{23} (V^{\dagger}y_{3L}^*)^{32} , (V^Ty_{3L})^{22} (V^{\dagger}y_{3L}^*)^{33}  \in [-0.040,0.040] \times \left( \frac{m_{S_3}}{\text{TeV}}\right)^2$ \\ \hline
 \multirow{3}{*}{$B \to K \nu \bar{\nu}$}  & $(V^Ty_{1L})^{3i} (V^{\dagger}y_{1L}^*)^{2j} \in [-0.070,0.029] \times \left( \frac{m_{S_1}}{\text{TeV}}\right)^2$ \\ 
  & $y_{2L}^{3i} y_{2L}^{*2j} \in [-0.032,0.061] \times \left( \frac{m_{R_2}}{\text{TeV}}\right)^2$ \\
  & $(V^Ty_{3L})^{3i} (V^{\dagger}y_{3L}^*)^{2j} \in [-0.070,0.029] \times \left( \frac{m_{S_3}}{\text{TeV}}\right)^2$ \\  \hline
  \multirow{3}{*}{$B_s^0-\overline{B_s^0}$}  & $(V^Ty_{1L})^{2i} (V^{\dagger}y_{1L}^*)^{3i} \in [-0.14,0.14] \times \left(\frac{m_{S_1}}{\text{TeV}}\right)^2$ \\ 
  & $y_{2L}^{2i} y_{2L}^{*3i} \in [-0.14,0.14] \times \left(\frac{m_{R_2}}{\text{TeV}}\right)^2$ \\
  & $(V^Ty_{3L})^{2i} (V^{\dagger}y_{3L}^*)^{3i} \in [-0.061,0.061] \times \left(\frac{m_{S_3}}{\text{TeV}}\right)^2$ \\ \hline 
  \multirow{3}{*}{$K^0-\overline{K^0}$}  & $(V^Ty_{1L})^{2i} (V^{\dagger}y_{1L}^*)^{1i} \in [-0.026,0.026] \times \left(\frac{m_{S_1}}{\text{TeV}}\right)^2$ \\ 
  & $y_{2L}^{2i} y_{2L}^{*1i} \in [-0.026,0.026] \times \left(\frac{m_{R_2}}{\text{TeV}}\right)^2$ \\
  & $(V^Ty_{3L})^{2i} (V^{\dagger}y_{3L}^*)^{1i} \in [-0.013,0.013] \times \left(\frac{m_{S_3}}{\text{TeV}}\right)^2$ \\ \hline 
 \end{tabular}
  \caption{Bounds on the leptoquark couplings from neutral mesons mixing and rare decay processes.}
 \label{mesonexp}
\end{table}

\subsection{Rare meson semi-leptonic decays}
The meson rare semi-leptonic decays can be induced at the tree-level by the leptoquarks and present constraints on the corresponding parameters. Here we consider $B \to K \nu \bar{\nu}$ and $B \to K^* \nu \bar{\nu}$ processes, related to the $(\bar qq\bar{\nu}\nu)$ interactions. The corresponding SM predictions are $\text{Br}(B^0 \to K^0 \nu \nu)=(4.1\pm 0.5) \times 10^{-6}$ and $\text{Br}(B^0 \to K^{*0} \nu \nu)=(9.2\pm 1.0) \times 10^{-6}$~\cite{Buras:2014fpa, Altmannshofer:2009ma}, while the current experimental upper limit bounds are given as $2.6\times 10^{-5}$ and $1.8 \times 10^{-5}$ by the Belle collaboration~\cite{Belle:2017oht} respectively. To describe the constraints on new physics from the $B \to K \nu \bar{\nu}$ and $B \to K^* \nu \bar{\nu}$ processes, the ratio $R_{K^{(*)}}^{\nu\nu}$ is introduced and defined as
\begin{equation}
R_{K^{(*)}}^{\nu\nu}=\frac{\text{Br}^{\text{SM+NP}}(B \to K^{(*)} \nu \bar{\nu})}{\text{Br}^{\text{SM}}(B \to K^{(*)} \nu \bar{\nu})}\,.
\end{equation}
The latest Belle results~\cite{Belle:2017oht} imply $R_{K}^{\nu\nu} < 3.9$ and $R_{K^{*}}^{\nu\nu}<2.7\,$.  As shown in Table~\ref{LQWilson}, the contributions to $B \to K^{(*)} \nu \bar{\nu}$ from leptoquarks $S_1$ and $S_3$ are represented by the Wilson coefficients $h_{V,d}^{LL}$, while by the Wilson coefficient $h_{V,d}^{RL}$ for the case of leptoquark $\tilde{R}_2$. If the new physics contribution is dominated by the $h_{V,d}^{LL}$ term, the ratios $R_{K}$ and $R_{K^{*}}$ can be calculated by the following formula~\cite{Buras:2014fpa},
\begin{equation}
R_{K^{(*)}}^{\nu\nu}=\frac{2}{3}+ \sum_{\nu, \nu^{\prime}} \frac{1}{3|C_{L}^{\text{SM}}|^2}\left| \delta^{\nu\nu^{\prime}} C_{L}^{\text{SM}}+(h_{V,d}^{LL})^{32;\nu\nu^{\prime}}\right|^2 \,,
\end{equation}
where $C_{L}^{\text{SM}}$ describes the SM contribution and the value is $C_{L}^{\text{SM}}=-6.35$. Note that since the experiments cannot detect the neutrinos in the final state, we need sum over all the flavor. On the other hand, if the new physics contribution is only originated from the $h_{V,d}^{RL}$ term, one has $R_{K}^{\nu\nu} \neq R_{K^{*}}^{\nu\nu}$ and the ratios are then presented by
\begin{align}
&R_{K}^{\nu\nu} =\frac{2}{3} + \sum_{\nu, \nu^{\prime}} \frac{1}{3|C_{L}^{\text{SM}}|^2}\left[\delta^{\nu\nu^{\prime}}C_{L}^{\text{SM}}+(h_{V,d}^{RL})^{32;\nu\nu^{\prime}}\right] \left[1+2 \frac{\delta^{\nu\nu^{\prime}}C_{L}^{\text{SM}}\,\text{Re}(h_{V,d}^{RL})^{32;\nu\nu^{\prime}}}{|C_{L}^{\text{SM}}|^2+|(h_{V,d}^{RL})^{32;\nu\nu^{\prime}}|^2} \right]\,, \\
&R_{K^*}^{\nu\nu} =\frac{2}{3} + \sum_{\nu, \nu^{\prime}} \frac{1}{3|C_{L}^{\text{SM}}|^2}\left[\delta^{\nu\nu^{\prime}}C_{L}^{\text{SM}}+(h_{V,d}^{RL})^{32;\nu\nu^{\prime}}\right] \left[1-1.34\frac{\delta^{\nu\nu^{\prime}}C_{L}^{\text{SM}}\,\text{Re}(h_{V,d}^{RL})^{32;\nu\nu^{\prime}}}{|C_{L}^{\text{SM}}|^2+|(h_{V,d}^{RL})^{32;\nu\nu^{\prime}}|^2} \right]\,.
\end{align}

\subsection{Neutral meson mixing}
Leptoquarks can induce neutral meson mixing via box diagrams mediated by leptons and leptoquarks. In this subsection we study the constraints from the $B_s^0-\overline{B_s^0}$ and $K^0-\overline{K^0}$ mixing. The related effective Hamiltonian can be described by~\cite{Bobeth:2017ecx}
\begin{equation}
\mathcal{H}_{\text{eff}}=C^{ij}_{LL}(\bar{d}_L^i\gamma^{\mu}d_L^j)(\bar{d}_L^i\gamma_{\mu}d_L^j)+C^{ij}_{RR}(\bar{d}_R^i\gamma^{\mu}d_R^j)(\bar{d}_R^i\gamma_{\mu}d_R^j)+C^{ij}_{LR}(\bar{d}_L^i\gamma^{\mu}d_L^j)(\bar{d}_R^i\gamma_{\mu}d_R^j)\,,
\end{equation}
where $i,j=3,2$ corresponding to $B_s^0-\overline{B_s^0}$ mixing and $i,j=2,1$ related to $K^0-\overline{K^0}$ mixing. Mapping the contribution of leptoquarks $S_1, \tilde{R}_2$ and $S_3$, we have the Wilson coefficients at the scale $\mu=m_{\text{LQ}}$ in the following form,
\begin{align}
&S_1: \qquad  C^{ij}_{LL}=-\frac{1}{128\pi^2 m_{S_1}^2} \sum_k \left[(V^T y_{1L})^{ik}(V^{\dagger}y_{1L}^*)^{jk}\right]^2 \,,\\
&\tilde{R}_2:  \qquad C^{ij}_{RR}=-\frac{1}{128\pi^2 m_{R_2}^2} \sum_k 2(y_{2L}^{ik} y_{2L}^{*jk})^2 \,,\\
&S_3: \qquad  C^{ij}_{LL}=-\frac{1}{128\pi^2 m_{S_3}^2} \sum_k 5\left[(V^T y_{3L})^{ik}(V^{\dagger}y_{3L}^*)^{jk}\right]^2\,.
\end{align}
The transition of the Wilson coefficients from $\mu=1\,\text{TeV}$ to $\mu=m_b$ scale are evaluated by the \textsl{Wilson} package~\cite{Aebischer:2018bkb} and the results are given by
\begin{equation}
C^{ij}_{LL,RR}(\mu=1\,\text{TeV})=0.78\,C^{ij}_{LL,RR}(\mu=m_b)\,.
\end{equation}

The current measurements of the mass differences in $B_s^0-\overline{B_s^0}$ and $K^0-\overline{K^0}$ mixing are~\cite{ParticleDataGroup:2020ssz},
\begin{align}
&\Delta m_{B_s}^{\text{exp}} = (17.741 \pm 0.020) \times 10^{12}\, \text{sec}^{-1}\,, \\
&\Delta m_{K}^{\text{exp}} = (3.484 \pm 0.0009) \times 10^{10}\, \text{sec}^{-1}\,.
\end{align}
For the mass difference $\Delta m_{B_s}^{\text{exp}}$, the SM prediction value is $\Delta m_{B_s}^{\text{SM}} = (18.3 \pm 2.7) \times 10^{12}\, \text{sec}^{-1}$~\cite{Artuso:2015swg, FermilabLattice:2016ipl, Jubb:2016mvq}. But the SM prediction for the mass difference in $K^0-\overline{K^0}$ mixing has not been precisely  estimated~\cite{Brod:2011ty, Buras:2014maa}. Thereby in our analysis,  we take the new physics contribution to $K^0-\overline{K^0}$ mixing to be compatible with the experimental value. The bounds on the leptoquarks couplings from neutral meson mixing and rare decay processes are summarized in Table~\ref{mesonexp}.

\section{Numerical analysis}
\begin{figure}[!t]
  \begin{minipage}{0.45\linewidth}
    \centerline{\includegraphics[width=1\textwidth]{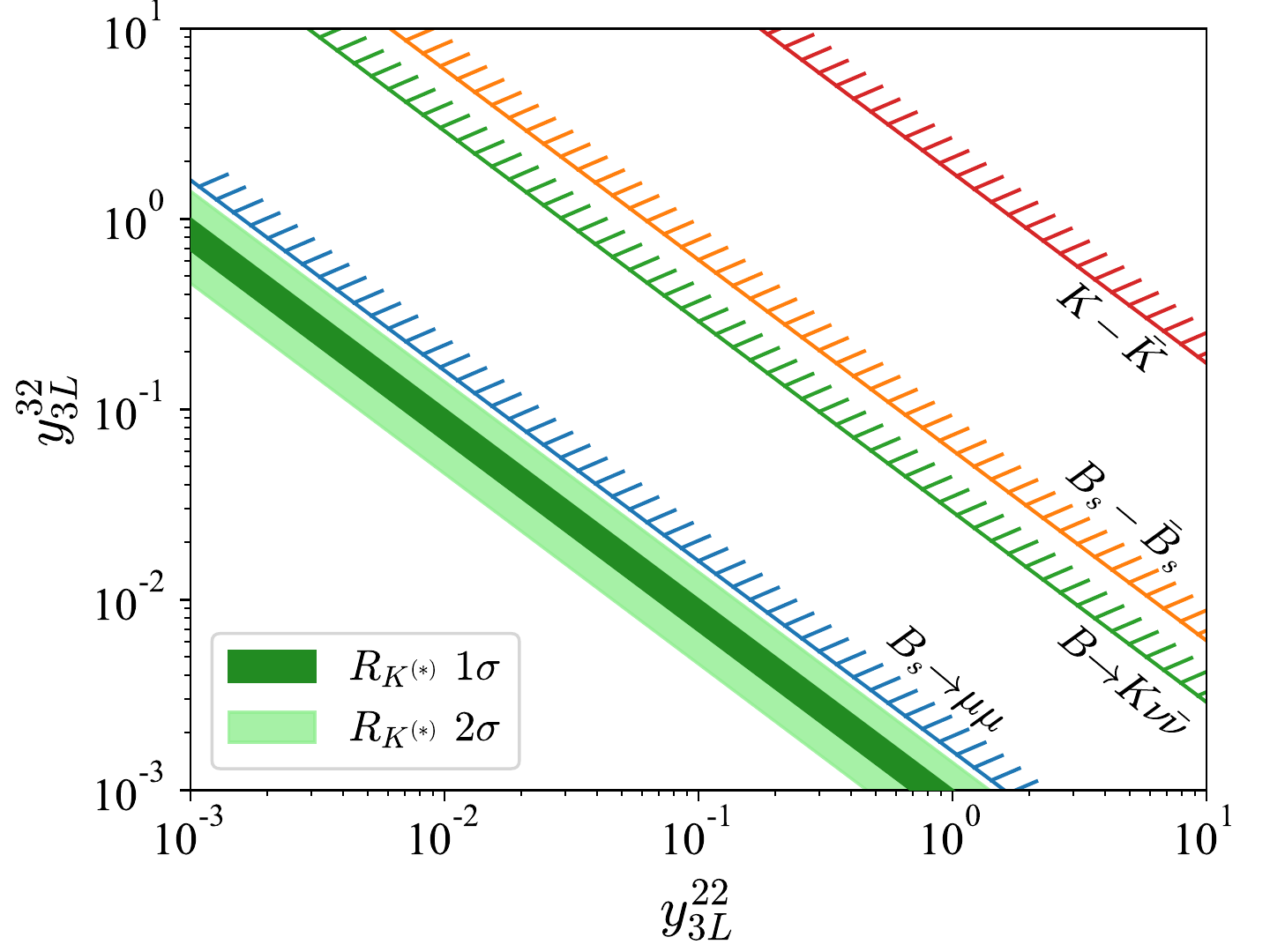}}
  \end{minipage}
  \begin{minipage}{0.45\linewidth}
    \centerline{\includegraphics[width=1\textwidth]{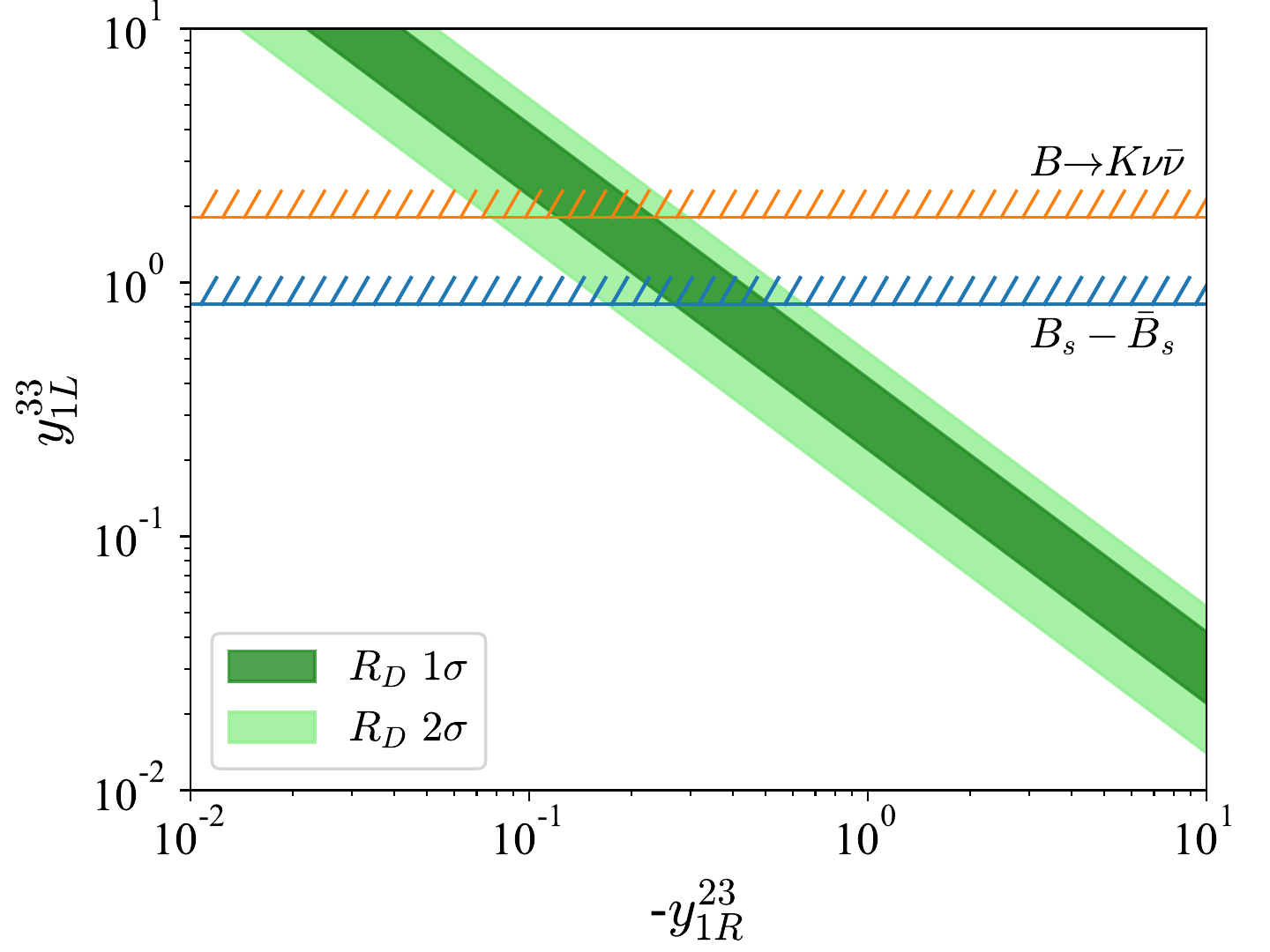}}
  \end{minipage} \\
  \begin{minipage}{0.45\linewidth}
    \centerline{\includegraphics[width=1\textwidth]{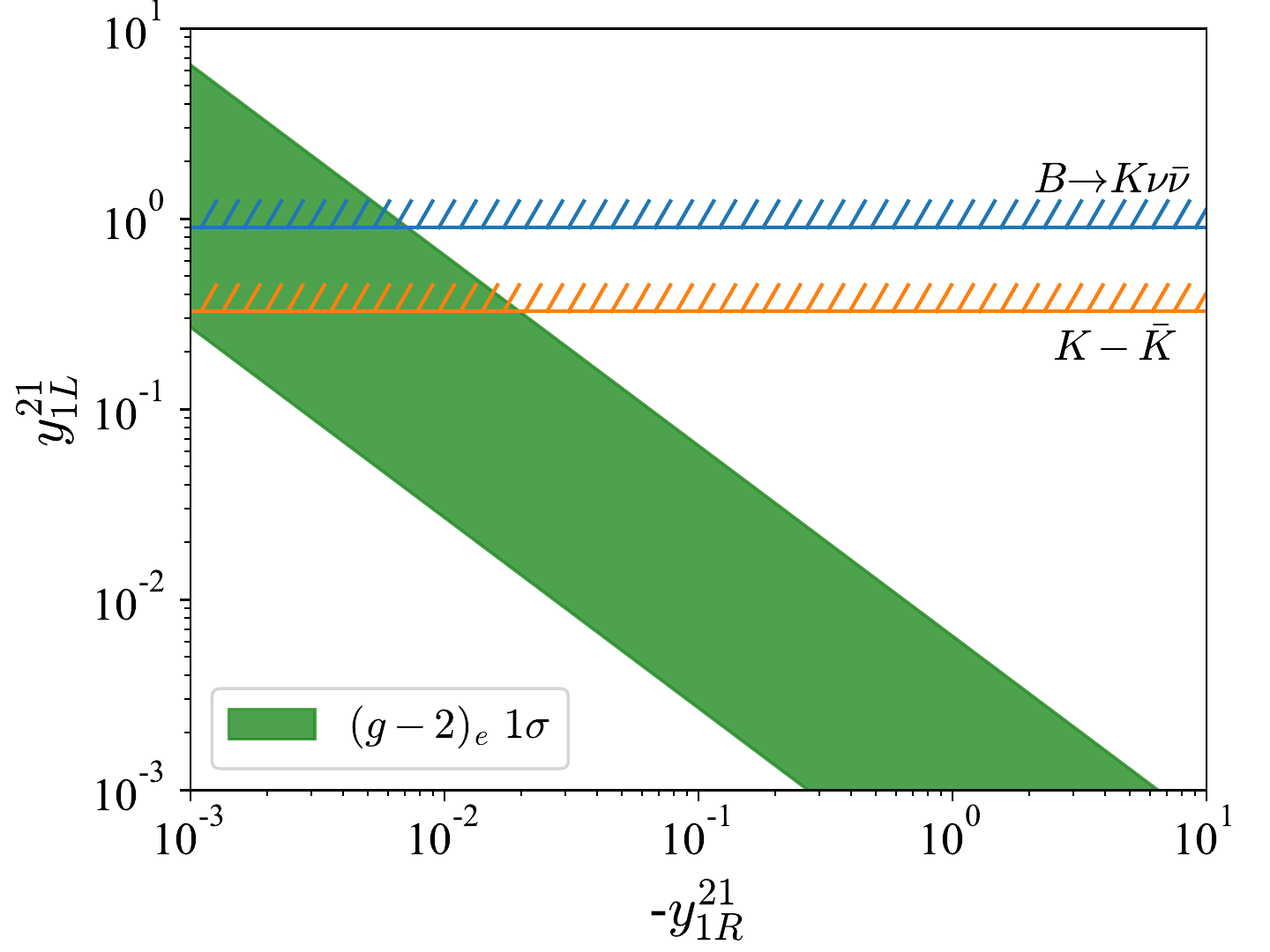}}
  \end{minipage}
  \begin{minipage}{0.45\linewidth}
    \centerline{\includegraphics[width=1\textwidth]{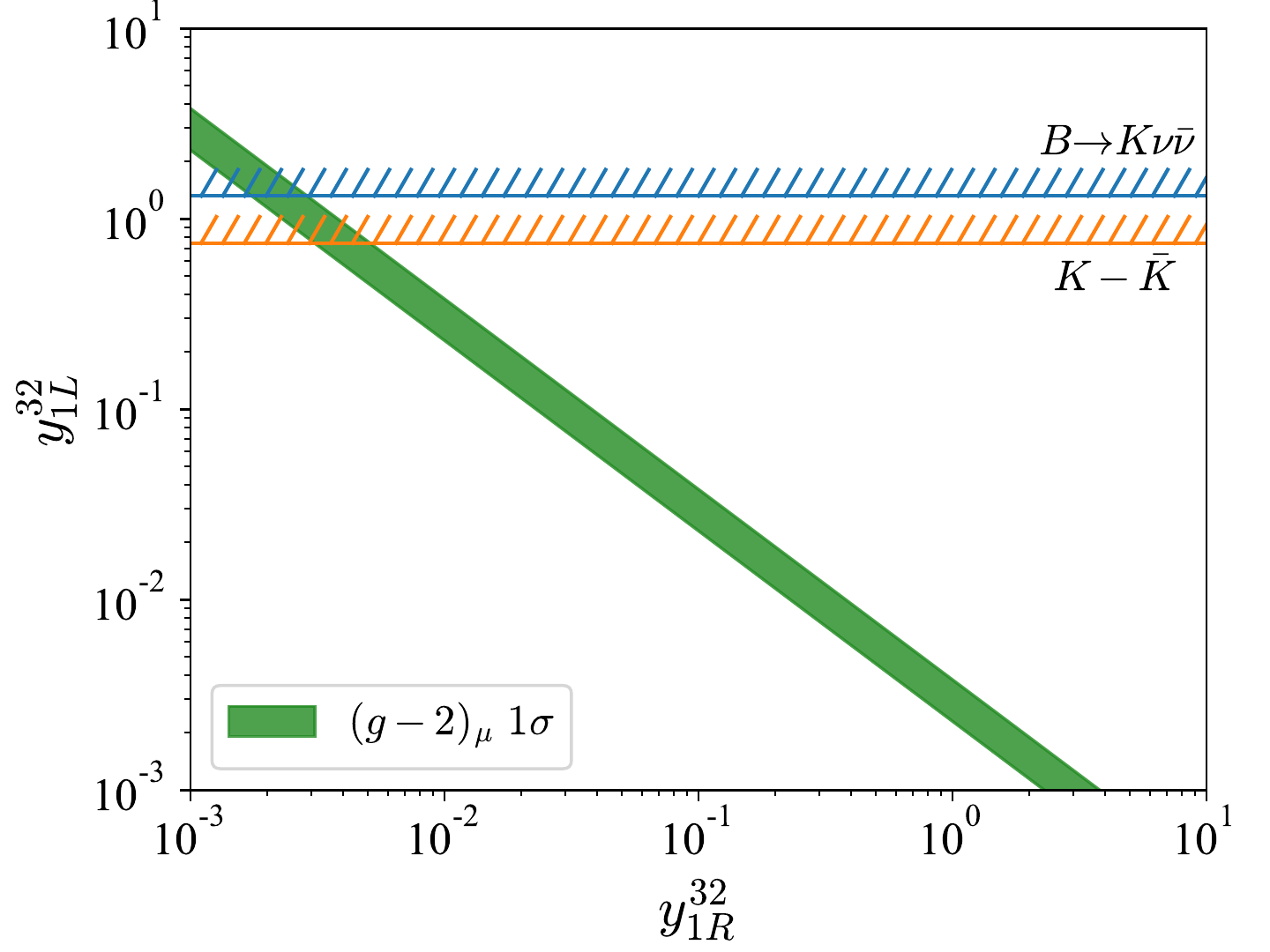}}
 \end{minipage}
\caption{Allowed regions of the various leptoquark Yukawa couplings obtained from fitting to the corresponding processes. The dark (light) green band represent the $1\sigma$ ($2\sigma$) allowed regions that explain the corresponding anomalies and the lines denote the constraints from the labelled processes.}
\label{para_sum}
\end{figure}
In this section, we perform a numerical analysis of  the model parameter space to supply a common explanation of $B$-physics anomalies in $R_{K^{(*)}}$,  $R_{D^{(*)}}$ and the charged leptons anomalous magnetic moment $(g-2)_{e,\mu}$, as well as the neutrino oscillation data. Instead of exploring the entire parameter space, we find the minimal parameters of the model and combine the constraints from the low-energy processes given in the Sec.~\ref{constraints}. We fix the components of singlet leptoquark $S_1$ and triplet leptoquark $S_3$ mass at 1 TeV ($m_{S_1}=m_{S_3}=1\,\text{TeV}$) and fix the components of doublet leptoquark $\tilde{R}_2$ mass at 2$\ \text{TeV}$ ($m_{R_2}=2\ \text{TeV}$). We use the python package Flavio to obtain the appropriate values of Wilson coefficients that explain the anomalies at the scale of leptoquark masses ($\mu=1\ \text{TeV}$) and then analyze the model parameter space.

In order to minimize the number of parameters, we adopt the following form of the Yukawa coupling matrices in the analysis.
\begin{align}
y _{1R} = \begin{pmatrix}
0\ & 0\ & 0 \\
\textcolor[RGB]{0,191,255}{y_{1R}^{21}} & 0\ & \textcolor[RGB]{255,165,0}{y_{1R}^{23}} \\
0\  & \textcolor[RGB]{0,128,0}{y_{1R}^{32}} & 0
\end{pmatrix}\,, &\qquad
y _{1L} = \begin{pmatrix}
0\ & 0\ & 0\ \\
\textcolor[RGB]{0,191,255}{y_{1L}^{21}} & 0\ & y_{1L}^{23}\\
0\ & \textcolor[RGB]{0,128,0}{y_{1L}^{32}} & \textcolor[RGB]{255,165,0}{y_{1L}^{33}}
\end{pmatrix}\,, \nonumber\\
y_{3L} = \begin{pmatrix}
0\ &  0  & 0\ \\
0\ & \textcolor[RGB]{255,69,0}{y_{3L}^{22}} & y_{3L}^{23}\ \\
y_{3L}^{31}\ & \textcolor[RGB]{255,69,0}{y_{3L}^{32}} & 0\
\end{pmatrix} \,, &\qquad
y_{2L} = \begin{pmatrix}
0\ & 0\ & 0\ \\
y_{2L}^{21}\ & y_{2L}^{22}\ & y_{2L}^{23} \\
y_{2L}^{31}\ & y_{2L}^{32}\ & y_{2L}^{33}
\end{pmatrix}\,.
\end{align}
The coupling combination $(y_{3L}^{22},\ y_{3L}^{32})$ can explain the anomalies of $R_K$ and $R_{K^*}$, while the couplings $(y_{2R}^{33},\ y_{2L}^{23})$ can contribute to $R_D$ and $R_{D^*}$. The couplings $(y_{1R}^{21},\ y_{1L}^{21})$ and $(y_{1R}^{32},\ y_{1L}^{32})$ give contributions to $\Delta a_e$ and $\Delta a_{\mu}$ respectively. The other non-zero couplings are needed to fit the neutrino masses and mixing angles. For a simple illustration, in Fig.~\ref{para_sum}, we present the allowed parameter space to explain these anomalies and satisfy the relevant processes constraints with taking the coupling as real. Specifically, we provide two concrete benchmark points of the leptoquarks Yukawa couplings. In benchmark point 1, the couplings are chosen as complex number. While for benchmark point 2, we choose the Dirac CP angle in the neutrino mixing matrix as $180^{\circ}$, which is within $1\, \sigma$ allowed range~\cite{ParticleDataGroup:2020ssz} and it is possible to take all the leptoquark Yukawa coupling values as real. The corresponding values of observables for these two benchmark points are summarized in Table~\ref{1trip_fit}.

\noindent Benchmark point 1:
\begin{align}
&y _{1R} = \begin{pmatrix}
0\,    & 0\,    & 0\, \\
-0.37\, & 0\,    & -0.70\, \\
0   \, & 0.054\, & 0\, 
\end{pmatrix}\,, \qquad
y_{3L} = \begin{pmatrix}
0\, &  0\,   & 0\, \\
0\, & 0.029\, & 0\, \\
0\, & 0.023\, & 0\,
\end{pmatrix}\,, \nonumber \\
&y _{1L} = \begin{pmatrix}
0\     & 0\,    & 0\, \\
0.012+0.016i\, & 0\,    & -0.049-0.0042i\, \\
0\,    & 0.57+0.0082i\, & 0.59+0.052i\,
\end{pmatrix}\,, \nonumber   \\
&y_{2L} = \begin{pmatrix}
0\,    & 0\,        & 0\, \\
0.043-0.042i \, & 0.044 \, & -0.048\, \\
-0.00013 \, & 0.00038 \, & 0.00027\,
\end{pmatrix}\,.
\label{BP1}
\end{align}
\noindent Benchmark point 2:
\begin{align}
&y _{1R} = \begin{pmatrix}
0\,    & 0\,    & 0\, \\
-0.014\, & 0\,    & -0.93\, \\
0   \, & 0.012\, & 0\,
\end{pmatrix}\,, \ \ \ \ 
y _{1L} = \begin{pmatrix}
0\     & 0\,    & 0\, \\
0.37\, & 0\,    & 0\, \\
0\,    & 0.21\, & 0.38\,
\end{pmatrix}\,,  \nonumber  \\
&y_{3L} = \begin{pmatrix}
0\, &  0\,   & 0\, \\
0\, & 0.070\, & 0.038\, \\
0.0019\, & 0.0059\, & 0\,
\end{pmatrix}\,, \ \ \ \ \ 
y_{2L} = \begin{pmatrix}
0\,    & 0\,        & 0\, \\
-0.015\, & -0.0020\,    & 0.023\, \\
0.0015\, & 0.0031\, & -0.00078\,
\end{pmatrix}\,.
\label{BP2}
\end{align}
\begin{table}[!ht]
\centering
 \begin{tabular}{|c|c|c|c|}\hline
Observables & Allowed range & BP1 & BP2 \\ \hline
 $\Delta m_{21}^2(10^{-5}\text{eV}^2)$ & [6.82, 8.04]         & 7.44 & 7.40  \\ \hline
 $\Delta m_{32}^2(10^{-3}\text{eV}^2)$ & [2.435, 2.598]       & 2.50 & 2.51 \\ \hline
 $\sin^2\theta_{12} $         & [0.269, 0.343]        & 0.305 & 0.301 \\ \hline
 $\sin^2\theta_{23} $         & [0.405, 0.620]        & 0.569 & 0.570 \\ \hline
 $\sin^2\theta_{13} $         & [0.02064, 0.02430]    & 0.0226& 0.0225 \\ \hline
 $\delta_{\text{CP}} /^{\circ}$ & [169, 246]          & 194   & 180     \\ \hline
 $R_{K}$                      & [0.795, 0.901]        & 0.808 & 0.812 \\ \hline 
 $R_{K^*}$                    & [0.569, 0.845]        & 0.794 & 0.817  \\ \hline 
 $R_{K_S^0}$                  & [0.48, 0.88]        & 0.808 & 0.812  \\ \hline 
 $R_{K^{*+}}$                 & [0.53, 0.91]        & 0.825 & 0.844  \\ \hline 
 $R_{D}$                      & [0.310, 0.370]        & 0.358 & 0.351  \\ \hline
 $R_{D^*}$                    & [0.281, 0.309]        & 0.305 & 0.292  \\ \hline
 $\Delta a_{e}\,(10^{-13})$   & [-12.3, -5.1]         & -8.48 & -9.89   \\ \hline
 $\Delta a_{\mu}\,(10^{-9})$  & [1.93, 3.11]          & 2.52  & 2.06  \\ \hline
 Br$(\mu \to e \gamma)$       & $<4.2\times 10^{-13}$ & $6.49\times 10^{-21}$ & $1.04\times 10^{-17}$  \\ \hline
 Br$(\tau \to e \gamma)$      & $<3.3\times 10^{-8}$  & $2.98\times 10^{-18}$ & $6.09\times 10^{-16}$  \\ \hline
 Br$(\tau \to \mu \gamma)$    & $<4.4\times 10^{-8}$  & $2.99\times 10^{-18}$ & $7.82\times 10^{-16}$  \\ \hline
 Br$(\mu - e)_{\text{Au}}$    & $<7\times 10^{-13}$   & $4.84\times 10^{-19}$ & $2.62\times 10^{-15}$  \\ \hline
 Br$(B^0_s \to \mu \mu)$      & [2.58, 3.28] $\times 10^{-9}$  & $2.93\times 10^{-9}$  &  $3.42\times 10^{-9}$   \\ \hline
 Br$(B^0_s \to \tau \tau)$    & $<6.8\times 10^{-3}$  & $7.82\times 10^{-7}$  &  $7.95\times 10^{-7}$   \\ \hline
 Br$(B^0_s \to \mu \tau)$     & $<1.4\times 10^{-5}$ & $2.11\times 10^{-14}$  &  $1.40\times 10^{-10}$   \\ \hline
 $R_{K}^{\nu\nu}$             & $<3.9$ & 1.3  &  0.75   \\ \hline
 $R_{K^*}^{\nu\nu}$           & $<2.7$ & 1.4  &  0.76  \\ \hline
 $\Delta m_{B_s}^{\text{SM+NP}}/\Delta m_{B_s}^{\text{SM}}$ & [0.85,1.15] & 1.03  &  1.01   \\ \hline
 $\Delta m_{K}^{\text{NP}}(10^{10}\,s^{-1})$ & $<0.95$ & 0.0016  &  0.52   \\ \hline
 \end{tabular}
  \caption{Summary of the observable values for the benchmark points.}
 \label{1trip_fit}
\end{table}
\section{Conclusion}
In this paper, we have proposed a simple model by extending SM with three TeV-scale scalar leptoquarks $S_1$, $\tilde{R}_2$ and $S_3$, where the source of tiny neutrino masses, the lepton flavor anomalies in $B$-meson decays  $(R_{K^{(*)}}$, $R_{D^{(*)}})$ and the tension in the charged lepton (electron and muon) anomalous magnetic moments have a common solution. In the model, $R_{K^{(*)}}$ anomalies are resolved by the leptoquark $S_3$ via the Wilson coefficients $C_{9, 10}^{\mu\mu}$. Leptoquark $S_1$ explains the anomalies of $R_{D^{(*)}}$ through the Wilson coefficients $g_{S_L}, g_{T}$, as well as the deviations of leptonic magnetic moments $(g-2)_{e,\mu}$ by one-loop level contribution. The small mixing of leptoquarks $S_1$ with $\tilde{R}_2$ or $\tilde{R}_2$ with $S_3$ can generate tiny neutrino masses. We analyze the parameter space of the leptoquark Yukawa couplings and obtain the corresponding viable region. We study the relevant experimental constraints and conclude there is an appropriate parameter space accommodate to combined explanation for these anomalies and deviations.

\noindent{\bf Acknowledgements.}
This work is supported in part by the National Science Foundation of China (12175082, 11775093) and the Fundamental Research Funds for the Central Universities (CCNU22LJ004).
\bibliographystyle{JHEP}
\bibliography{references}
\end{document}